\pdfoutput=1
\documentclass[iop]{emulateapj}
\usepackage{natbib}
\usepackage{graphicx}

\received{receipt date}
\revised{revision date}
\accepted{acceptance date}

\shorttitle{A Multi-Wavelength High Resolution Study of S255}
\shortauthors{Zinchenko, Liu, Su, et al.}
\def\htwo{H\,{\scriptsize\sc{II}}}

\begin{document}

\title{A Multi-Wavelength High Resolution Study of the S255 Star Forming Region.\\ General structure and kinematics.}
\author{I. Zinchenko}
\affil{Institute of Applied Physics of the Russian Academy of Sciences, 46 Ulyanov st., Nizhny Novgorod 603950, Russia}
\affil{Nizhny Novgorod University, 23 Gagarin av., Nizhny Novgorod 603950, Russia }
\email{zin@appl.sci-nnov.ru}
\author{S.-Y. Liu, Y.-N. Su}
\affil{Institute of Astronomy and Astrophysics, Academia Sinica.
P.O. Box 23-141, Taipei 10617, Taiwan, R.O.C. }
\author{S. Kurtz}
\affil{Centro de Radioastronom\'\i a y Astrof\'\i sica, Universidad Nacional
Aut\'onoma de M\'exico, Morelia, Michoac\'an, M\'exico}
\author{D. K. Ojha}
\affil{Infrared Astronomy Group, Department of Astronomy and Astrophysics,
  Tata Institute  of  Fundamental Research, Homi Bhabha Road, Colaba,
  Mumbai (Bombay) -- 400 005, India}
\author{M. R. Samal}
\affil{Laboratoire d'Astrophysique de Marseille (UMR 6110 CNRS \& Universit\'e de Provence), 38 rue F. Joliot-Curie, 13388 Marseille Cedex 13, France}
\author{S. K. Ghosh}
\affil{National Centre for Radio Astrophysics, Tata Institute of Fundamental Research, Pune 411007, India}

\begin{abstract}
We present observational data for two main components (S255IR and S255N) of the S255 high mass star forming region in continuum and molecular lines obtained at 1.3 mm and 1.1 mm with the SMA, at 1.3 cm with the VLA and at 23 and 50 cm with the GMRT. The angular resolution was from $ \sim 2'' $ to $ \sim 5'' $ for all instruments. With the SMA we detected a total of about 50 spectral lines of 20 different molecules (including isotopologues). About half of the lines and half of the species (in particular N$_2$H$^+$, SiO, C$^{34}$S, DCN, DNC, DCO$^+$, HC$_3$N, H$_2$CO, H$_2$CS, SO$_2$) have not been previously reported in S255IR and partly in S255N at high angular resolution. Our data reveal several new clumps in the S255IR and S255N areas by their millimeter wave continuum emission. Masses of these clumps are estimated at a few solar masses. The line widths greatly exceed expected thermal widths. These clumps have practically no association with NIR or radio continuum sources, implying a very early stage of evolution. At the same time, our SiO data indicate the presence of high-velocity outflows related to some of these clumps. In some cases, strong molecular emission at velocities of the quiescent gas has no detectable counterpart in the continuum. We discuss the main features of the distribution of NH$_3$, N$_2$H$^+$, and deuterated molecules. We estimate properties of decimeter wave radio continuum sources and their relationship with the molecular material.
\end{abstract}
\keywords{astrochemistry -- HII regions -- instrumentation: interferometers -- ISM: clouds -- ISM: molecules -- stars: formation}

\section{Introduction}
Despite their great importance in almost all areas of astronomy, the formation of stars larger than 8--10 M$_\odot$ is still poorly understood. In part this is because high mass star formation (HMSF) regions are more distant, more active, and shorter lived than their low-mass counterparts. Nevertheless, an enhanced attention has been paid recently to the earliest phases of massive star formation. So far, few young HMSF cores have been studied in detail and it is important to extend the list of such objects, especially considering the large variety of processes that probably lead to the formation of a massive star. 

S255 is an \htwo\ region associated with a dense core at a commonly accepted distance of 2.5~kpc \citep{Russeil07,Ojha11} which we adopt here too. We note, however, that \citet{Rygl10} report a distance of 1.6~kpc based on trigonometric parallax measurements of methanol masers.  The core consists of two main components (S255IR and S255N) separated by slightly over $1'$. Our team has previously acquired molecular line observations using single-dish instruments (OSO-20m, IRAM-30m, NRAO-12m).  With angular resolution from about $1'$ to $10^{\prime\prime}$ \citep{Zin09}, we obtained $M \sim 300$~M$_{\odot}$, $n \sim 2 \cdot 10^5$~cm$^{-3}$, $T_{\mathrm{kin}} \sim 40$~K and $\Delta V \sim 2$~km\,s$^{-1}$ for both components. While both components show evidence for cool, massive clumps, their evolutionary states appear to be quite different. S255IR is bright ($> 70$ Jy) in all the Midcourse Space Experiment (MSX) bands and contains a near-IR cluster of early-B-type stars \citep{Howard97,Itoh01}, a cluster of compact \htwo\ regions \citep{Snell86}, and a wealth of complex
H$_2$ emission features \citep{Miralles97}. In contrast, S255N (also called Sh2-255 FIR1 and G192.60-MM1) contains a single cometary UC \htwo\ region \citep[e.g.,][]{Kurtz94}, and is undetected by MSX at wavelengths shorter than 21~$\mu$m \citep{Crowther03}. VLA and SMA observations in the continuum, several molecular lines and water maser emission \citep{Cyganowski07} indicate the presence of a massive protocluster in this region. The chemical composition of S255IR and S255N also appear significantly different \citep{Lintott05,Zin09}. While the CS and HCN abundances are very similar, the abundances of NH$_3$, N$_2$H$^+$, HCO$^+$ and some other molecules in these components differ significantly.  

Recently, both components were studied with the SMA at 1.3 mm in CO, $^{13}$CO, C$^{18}$O, CH$_3$OH, CH$_3$CN and some other lines \citep{Wang11}. These observations revealed 3 continuum clumps in the S255IR area and high-velocity collimated outflows in both regions. The star population in this complex was studied most recently by \citet{Ojha11} on the basis of optical and NIR observations. They found a number of new YSO candidates with a large spread in ages, indicating a scenario of induced star formation.

In general, this star forming complex represents an excellent laboratory for studies of different stages of massive star formation. Our goal is to investigate further the structure, physical properties and chemistry of this area on small scales by observations of important molecular tracers like NH$_3$ (which is a convenient ``thermometer" for
dense molecular cores and provides information on their density and kinematics too --- e.g. \citealt{Ho83,Walmsley83}), N$_2$H$^+$, SiO, deuterated molecules, etc. In addition we want to investigate in more detail the distribution and properties of ionized gas in this region in order to trace the interaction of massive YSOs and outflows with the surrounding medium. 

In this paper we present observational data obtained at 1.3 mm and 1.1 mm with the SMA in the compact configuration (both in various lines and in continuum), with the VLA at 1.3 cm (in ammonia lines) and with the GMRT at 23 and 50 cm in continuum. We also discuss the general structure and kinematics of the complex and derive basic physical parameters of the observed features.

\section{Observations and data reduction}

\subsection{SMA}
The S255 complex, including S255~N and S255~IR, was observed with the Submillimeter Array (SMA) in its compact
configuration on 2010 April 7th and April 14th at 224.8~GHz and 283.9~GHz, respectively.
Two fields, one centered at S255~N (06$^h$12$^m$53$^s$.669, 18$^\circ$00$'$26$''$.903)
and one centered at S255~IR (06$^h$12$^m$53$^s$.800, 17$^\circ$59$'$22$''$.097) were interleaved throughout the observations. The primary HPBW of the SMA antennas is $55''-45''$ at these frequencies.
Typical system temperatures on source were between 90~K and 160~K for the 224.8~GHz track,
and between 100~K and 200~K for the 283.9~GHz track.
The resulting $uv$-coverage ranges from 6~k$\lambda$ to 50~k$\lambda$ and 8~k$\lambda$ to 65~k$\lambda$ for the 224.8 GHz and 283.9~GHz tracks, respectively.
3C273 was used as the bandpass calibrator for both tracks,
and 0530+135 as well as 0750+125 were used as the complex phase and amplitude gain calibrator.
The gain calibrator flux scale, calibrated against Mars, was found to be consistent within 5\% with the SMA calibrator database
and estimated to be accurate within 20\%.
In the 224.8 GHz observation, a total of 8~GHz (216.8$-$220.8~GHz in the LSB and 228.8$-$232.8~GHz in the USB)
was observed with the SMA bandwidth doubling correlator configuration.
A number of different spectral resolutions, with a maximum resolution of 0.53 km~s$^{-1}$, were employed for different molecular lines.
The 283.9 GHz observation covered 277.9$-$279.9~GHz in the LSB and 287.9$-$289.9 GHz in the USB.
The data calibration was carried out with the IDL superset MIR \citep{Scoville93},
and subsequent imaging and analysis were done in MIRIAD \citep{Sault95}.
With robust weighting for the continuum and line data,
the synthesized beam sizes are approximately $3\farcs8 \times 3\farcs0$ (PA 60$^\circ$) at 1.3~mm and 
$2\farcs9 \times 2\farcs6$ (PA 40$^\circ$) at 1.1~mm.
The noise level is $\sim$ 8 mJy beam$^{-1}$ in the 1.1 mm continuum image
and roughly 60~mJy\,beam$^{-1}$ and 100~mJy\,beam$^{-1}$ in the line cubes at 2.0 km\,s$^{-1}$ spectral resolution at 220--230~GHz and 280--290~GHz, respectively.

\subsection{VLA}
The VLA observations (program AZ186) were made on 2009 
November 13 (S255N) and 21 (S255IR)  with the array in the D-configuration.
The pointing centers for S255N and S255IR in J2000
coordinates were 06$^h$12$^m$53$^s$.7 +18$^\circ$00$'$27$''$ and 06$^h$12$^m$54$^s$.0
+17$^\circ$59$'$22$''$, respectively.   At the observing frequency the VLA primary beam is
approximately 2 arcmin.  The correlator was in the 1IF mode,
measuring right circular polarization with a 6.25 MHz bandwidth
and 127 channels of 48.8 kHz each. The ammonia (1,1) and (2,2) lines were observed, with rest
frequencies of 23,693.7955 and 23,721.9336 MHz.
3C147 was observed as the flux calibrator, and J0530+135 was
observed as the phase and bandpass calibrator. The assumed flux density for 3C147 was 1.700 Jy and the
bootstrapped flux density for J0530+135 was $1.439\pm 0.015$~Jy.
The angular resolution of the image cubes was $2\farcs61 \times 2\farcs51$
(PA $-65.6^\circ$). The approximate rms noise
of the cube was 3.5 mJy~beam$^{-1}$.  The on-source time was approximately
one half hour in each transition. 

\subsection{GMRT}
The radio continuum interferometric observations at 610 MHz and 1280 MHz were carried out on 2009 May 08 and 2009 September 08, 
using the Giant Metrewave Radio Telescope (GMRT) array. The GMRT has a ``Y''-shaped hybrid configuration of 30 antennas, each 45 m in diameter. There are six
antennas along each of the three arms (with arm length of $\sim 14$~km).
The remaining twelve antennas are located in a random and compact
($\sim 1 {\rm km} \times 1 {\rm km}$) arrangement near the center. Details of the GMRT antennas and their configurations can be found in \citet{Swarup91}. The largest angular scales to which the GMRT is sensitive are 8$'$ and 17$'$ at 1280 and 610 MHz, respectively.

Flux and phase calibrators were observed to derive the phase and amplitude gains of the antennas. 3C147 was used as the flux calibrator, while the phase calibrators were 0503+020 and 0632+103 at 610 MHz and 1280 MHz, respectively. The data analysis was done using AIPS. Corrupted data were identified and flagged. The estimated uncertainty of the flux calibration is within 7\% at both frequencies.
Images of the field were formed by the Fourier inversion and cleaning algorithm task IMAGR. Several iterations of self-calibration were carried out to remove the residual effects of atmospheric and ionospheric phase corruption and to improve the quality of the maps. The high resolution images ($6\farcs4 \times 4\farcs8$ at 610~MHz and $5\farcs2 \times 3\farcs8$ at 1280~MHz) were made with a Briggs weighting function halfway between uniform and natural weighting (robust factor = --0.5 to 0), which is a good compromise between angular resolution and sensitivity. 

\section{Results}
With the SMA we detected about 50 spectral lines of 20 different molecules (including isotopologues). About half of the lines and half of the species (in particular $^{12}$CO, $^{13}$CO, C$^{18}$O, CH$_3$OH, CH$_3$CN, $^{13}$CS, HNCO, SO, OCS) were previously observed in this area with the SMA by \citet{Wang11}. We do not discuss these lines in detail; rather, we compare their results with our line detections. Our additional to \citet{Wang11} observations include lines of N$_2$H$^+$, SiO, C$^{34}$S, DCN, DNC, DCO$^+$, HC$_3$N, H$_2$CO, H$_2$CS, SO$_2$ as well as additional transitions of CH$_3$OH, HNCO and OCS. A list of these lines, including their frequencies and lower energy levels, is given in Table~\ref{table:lines}. Some of them (marked by asterisk in Table~\ref{table:lines}) were observed earlier in S255N at the SMA by \citet{Cyganowski07}. Nevertheless, they have not been observed in S255IR and our data show many new features in these lines in S255N, which are presented and discussed below. In addition, we observed the (1,1) and (2,2) ammonia lines with the VLA. Here we particularly focus on several key species, including N$_2$H$^+$, NH$_3$, deuterated molecules, SiO, methanol, etc. It is worth noting that we did not detect some important lines in our band, in particular, N$_2$D$^+$ and radio recombination lines.

We present the results in the form of maps as well as spectra and line parameters at selected positions. For continuum observations we give positions, flux densities, and size estimates of the continuum sources. In general the results are presented separately for the S255IR and S255N areas.

\begin{deluxetable}{lcrrc}
\tablecaption{List of molecular transitions observed at the SMA in either S255IR or S255N in addition to those observed by \citet{Wang11}. \label{table:lines}}
\tablehead{\colhead{Molecule} &\colhead{Transition} &\colhead{Frequency} &\colhead{$E_l$} & \\ \colhead{} &\colhead{} &\colhead{(GHz)} &\colhead{(cm$ ^{-1} $)} & }
\tablecolumns{5}
\startdata
CH$ _{3} $OH	&$5_{1} - 4_{2}$ E	&216.945521	&31.596	\\
				&$6_{1} - 7_{2}$ A$^-$	&217.299205	&252.644 \\
&$4_{2} - 3_{1}$	E	&218.440050		&24.310	&*\\
&$10_{2} - 9_{3}$ A$^+$	&232.418571		&107.208\\
&$9_{-1} - 8_{0}$ E		&278.304575		&67.150	\\
&$14_{4} - 15_{3}$ E	&278.599037		&226.775\\
&$11_{2} - 10_{3}$ A$^-$	&279.351887		&123.338\\
&$6_{0} - 5_{0}$	E	&289.939386		&33.316	\\
SiO		&5--4	&217.104984		&14.484	&*\\
C$^{34}$S	&6--5	&289.209068		&24.119	\\
DCN		&3--2	&217.238530		&7.246	&*\\
		&4--3	&289.644907		&14.493	\\
DNC		&3--2	&228.910471		&7.636	\\
DCO$^+$	&4--3	&288.143858		&14.418	\\
HNCO		&$ 10_{1,10}-9_{1,9} $	&218.981170		&62.949	\\
HC$_3$N	&24--23	&218.324711		&83.755	&*\\
H$_2$CS	&$8_{1,7} -    7_{1,6}$	&278.887680		&41.733	\\
H$_2$CO	&$3_{0,3} -    2_{0,2}$	&218.222195		&7.286	&*\\
		&$3_{2,2} -    2_{2,1}$	&218.475642		&40.040	&*\\
		&$3_{2,1} -    2_{2,0}$	&218.760071		&40.043	&*\\
N$_2$H$^+$	&3--2	&279.511760		&9.324	\\
OCS		&18--17	&218.903357		&62.070	\\
		&23--22	&279.685318		&102.634	\\
SO$_2$		&18$_{1,17}-17_{2,16}$ 	&288.519996		&103.712\\ 
\enddata
\tablenotetext{*}{Observed earlier at the SMA in S255N by \citet{Cyganowski07}}

\end{deluxetable}

\subsection{Millimeter wave continuum}

\begin{figure*}
\begin{minipage}{0.48\textwidth}
\centering
\includegraphics[width=\textwidth]{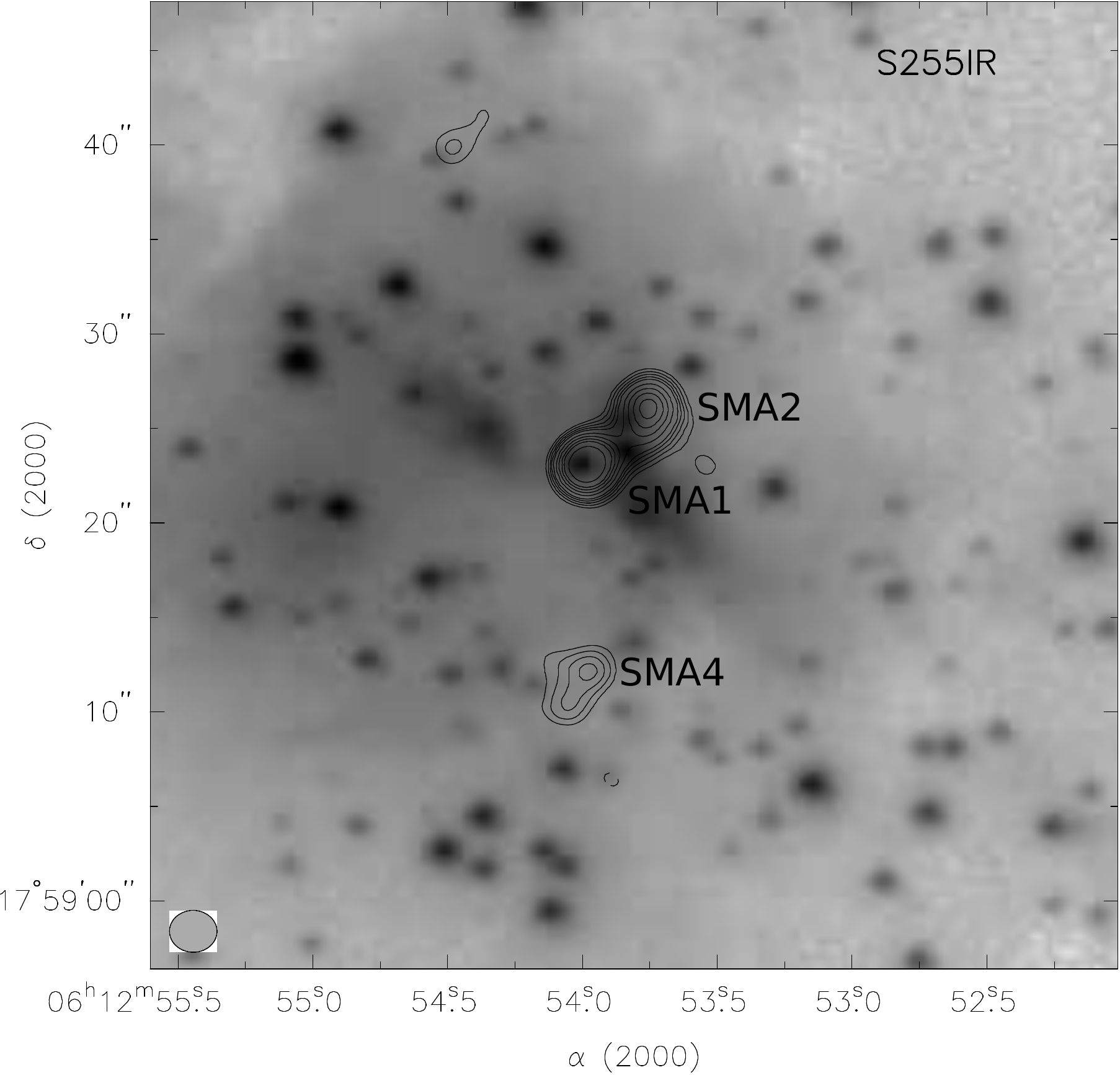}
\end{minipage}
\hfill
\begin{minipage}{0.48\textwidth}
\centering
\includegraphics[width=\textwidth]{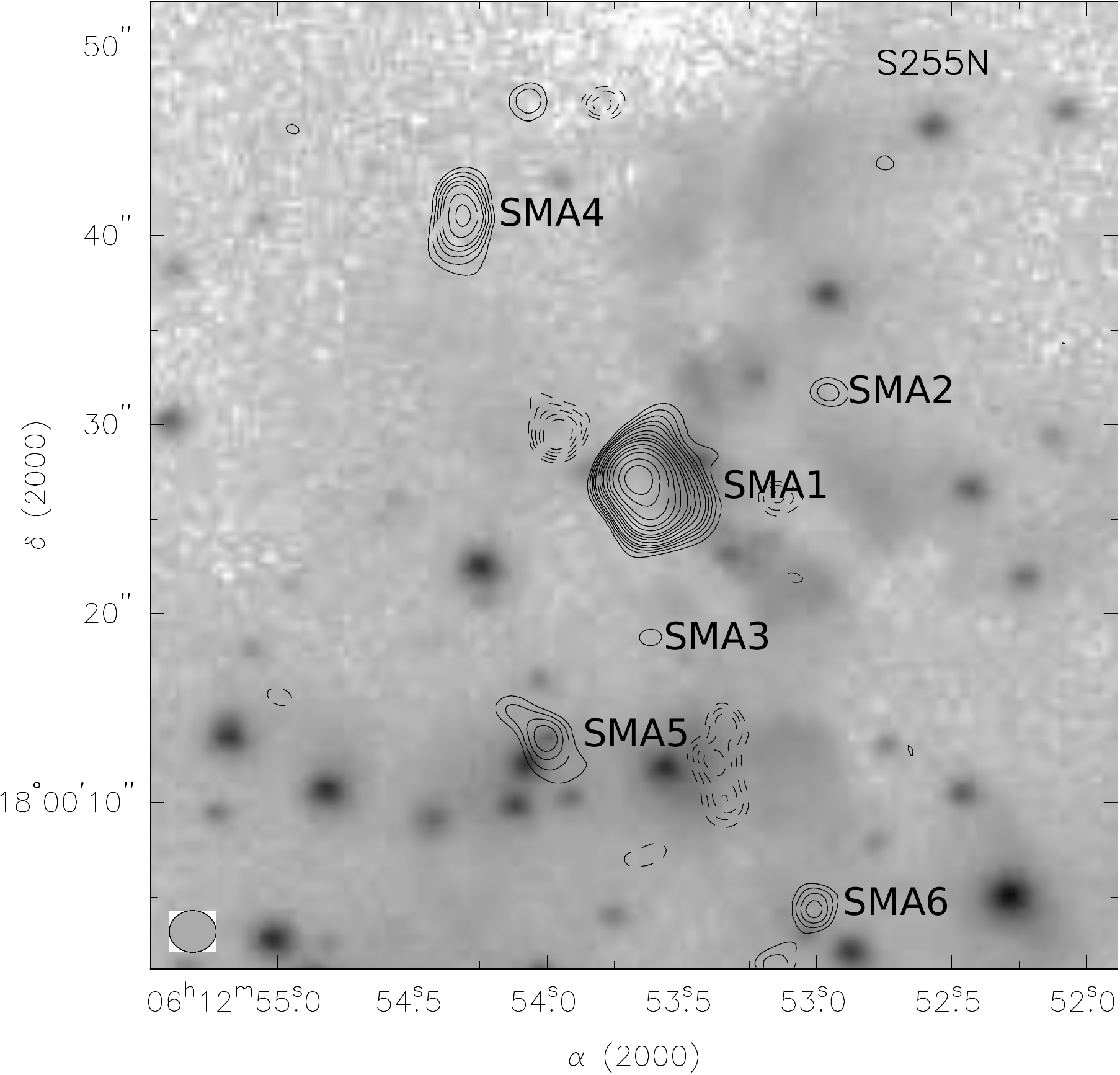}
\end{minipage}
\caption{Maps of the 1.1 mm continuum emission in the S255IR (left panel) and S255N (right panel) areas (contours) overlaid on the K band NIR images (in logarithmic scale) from \citet{Ojha11}. The contour levels are (2, 2.5, 3, 3.5, 4, 5, 6, 7, 8, 10)$\times 25$~mJy\,beam$^{-1}$ for S255IR and (--4, --3.5, --3, --2.5, --2, 2, 2.5, 3, 3.5, 4, 5, 6, 7, 8, 10, 12, 15, 20, 25)$\times 12.5$~mJy\,beam$^{-1}$ for S255N. The dashed contours show negative features due to the missing flux. The SMA beam is shown in the lower left corner of both panels.}
\label{fig:NIR}
\end{figure*}

Maps of the S255IR and S255N areas in 1.1~mm continuum overlaid on NIR K-band images \citep{Ojha11} are shown in Fig.~\ref{fig:NIR}. The two brightest continuum sources visible in the S255IR map have been reported and discussed by \citet{Wang11} (S255IR-SMA1 and S255IR-SMA2 in their nomenclature). In addition, they identified the S255IR-SMA3 clump which is indistinguishable from S255IR-SMA1 in our data (due to our lower angular resolution). On the other hand, our map shows an additional, very bright clump about 10$''$ further south. We designate this clump as S255IR-SMA4. 

In the S255N area, in addition to the continuum clumps reported earlier \citep{Cyganowski07,Wang11} our 1.1~mm continuum map shows three new features that are also detected in molecular emission (see below). We designate these as S255N-SMA4, SMA5 and SMA6; their parameters are given in Table~\ref{table:cont}.

\begin{deluxetable*}{lllcccrc}
\tablecaption{Positions, flux densities, deconvolved angular sizes and position angles of the millimeter wave continuum sources at 284~GHz as well as flux densities at 225~GHz (the last column) measured with the SMA. \label{table:cont}}
\tablehead{\colhead{Name} &\colhead{$\alpha$(2000)} &\colhead{$\delta$(2000)} &\colhead{$S_{284}$} &\colhead{$\theta_{\mathrm{max}}$} &\colhead{$\theta_{\mathrm{min}}$} &\colhead{P.A.} &\colhead{$S_{225}$}\\ \colhead{} &\colhead{(h m s)} &\colhead{($^\circ$ $^\prime$ $^{\prime\prime}$)} &\colhead{(Jy)} &\colhead{$^{\prime\prime}$} &\colhead{$^{\prime\prime}$} &\colhead{($^\circ$)} &\colhead{(Jy)}}
\tablecolumns{8}
\startdata
S255IR-SMA1 &6:12:53.98 &17:59:23.1 &0.45 &1.5    &0.7 &--14 &0.29\\
S255IR-SMA2 &6:12:53.75 &17:59:25.7 &0.43 &2.9    &2.2 &11 &0.30\\
S255IR-SMA4 &6:12:54.03 &17:59:11.5 &0.31 &4.4    &3.5 &--50 &0.08\\ 
S255N-SMA1  &6:12:53.64 &18:00:26.8 &0.75 &3.0    &2.0 &--4 &0.54\\
S255N-SMA2  &6:12:52.95 &18:00:31.7 &0.05 &2.5    &2.1 &--10 &0.06\\
S255N-SMA3  &6:12:53.62 &18:00:18.7 &0.05 &3.3    &2.9 &--31 &$ <0.03 $\\
S255N-SMA4  &6:12:54.32 &18:00:41.0 &0.16 &3.8    &0.7 &--12 &0.04\\
S255N-SMA5  &6:12:54.00 &18:00:13.2 &0.15 &4.5    &3.5 &49 &0.04\\
S255N-SMA6  &6:12:53.04 &18:00:03.4 &0.14 &7.9    &1.3 &--40 &$ <0.03 $

\enddata

\end{deluxetable*}

\subsection{Molecular emission}

Our data show that the intensity distributions for many molecular species are quite different from the continuum maps. In many cases, the molecular emission peaks do not coincide with the continuum peaks and there is also molecular emission without corresponding continuum emission. Below we describe in more detail the emission features of several key species.

\subsubsection{NH$_3$} \label{sec:nh3}
\begin{figure*}
\includegraphics[width=\textwidth]{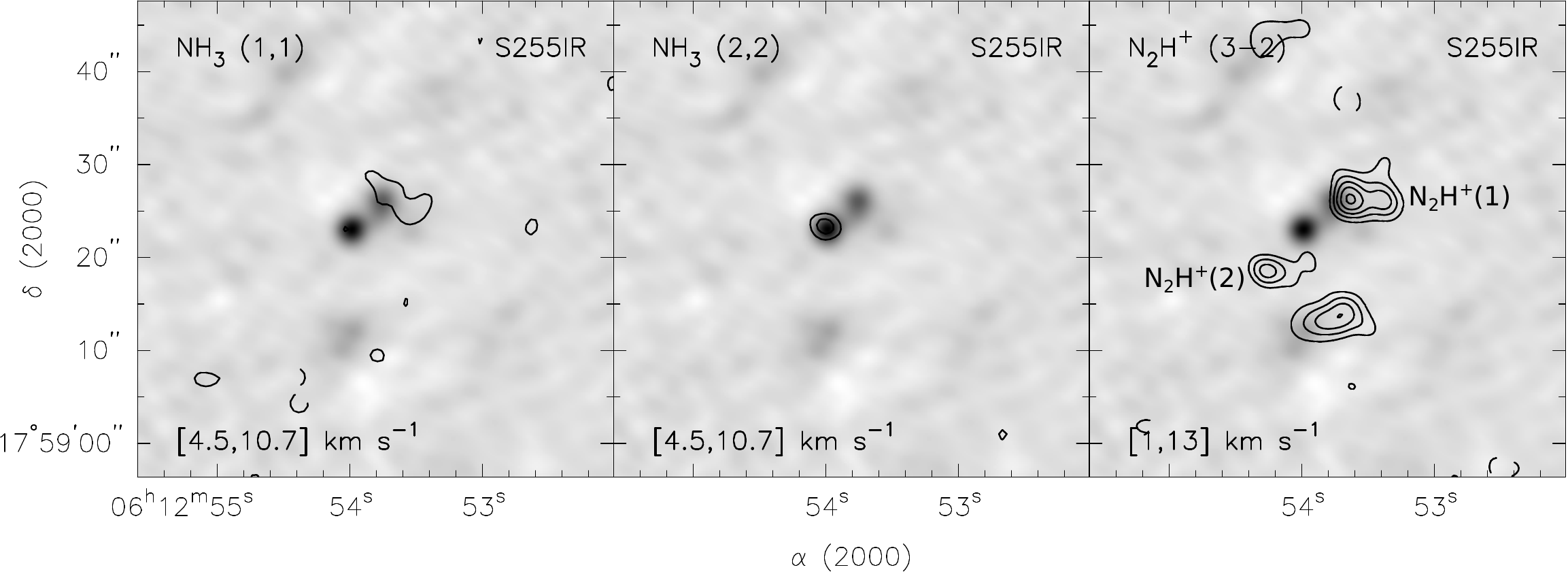} 
\caption{Maps of the NH$_3$ (1,1), NH$_3$ (2,2) and N$_2$H$^+$ (3--2) integrated line emission in the S255IR area (contours) overlaid on the 1.1~mm continuum image. The velocity ranges are indicated at the bottom of the plots. The contour levels are (--3, 3, 5, 7)$\times 10$~mJy\,beam$^{-1}$\,km\,s$^{-1}$ for NH$_3$ and (--3, 3, 5, 7, 9, 11)$\times 1.4$~Jy\,beam$^{-1}$\,km\,s$^{-1}$ for N$_2$H$^+$. The dashed contours show negative features due to the missing flux.}\label{fig:ir_fig1} 
\end{figure*}

The ammonia emission in the S255IR area is rather weak (in accordance with \citealt{Zin97}) and shows significantly different distributions in the NH$_3$ (1,1) and (2,2) transitions (Fig.~\ref{fig:ir_fig1}). The (1,1) emission peaks at the SMA2 clump while the (2,2) emission peak coincides with SMA1. This indicates that the temperature of the SMA1 clump is much higher than that of SMA2. Indeed, \citet{Wang11} derived a kinetic temperature of 150~K for SMA1, based on CH$_3$CN data, implying the presence of a hot core. The main emission peak towards SMA1 is observed at $ V_{\mathrm{LSR}}\approx 4.8 $~km\,s$^{-1}$. There is a hint on another peak at $ V_{\mathrm{LSR}}\approx 9 $~km\,s$^{-1}$.

\begin{figure*}
\includegraphics[width=\textwidth]{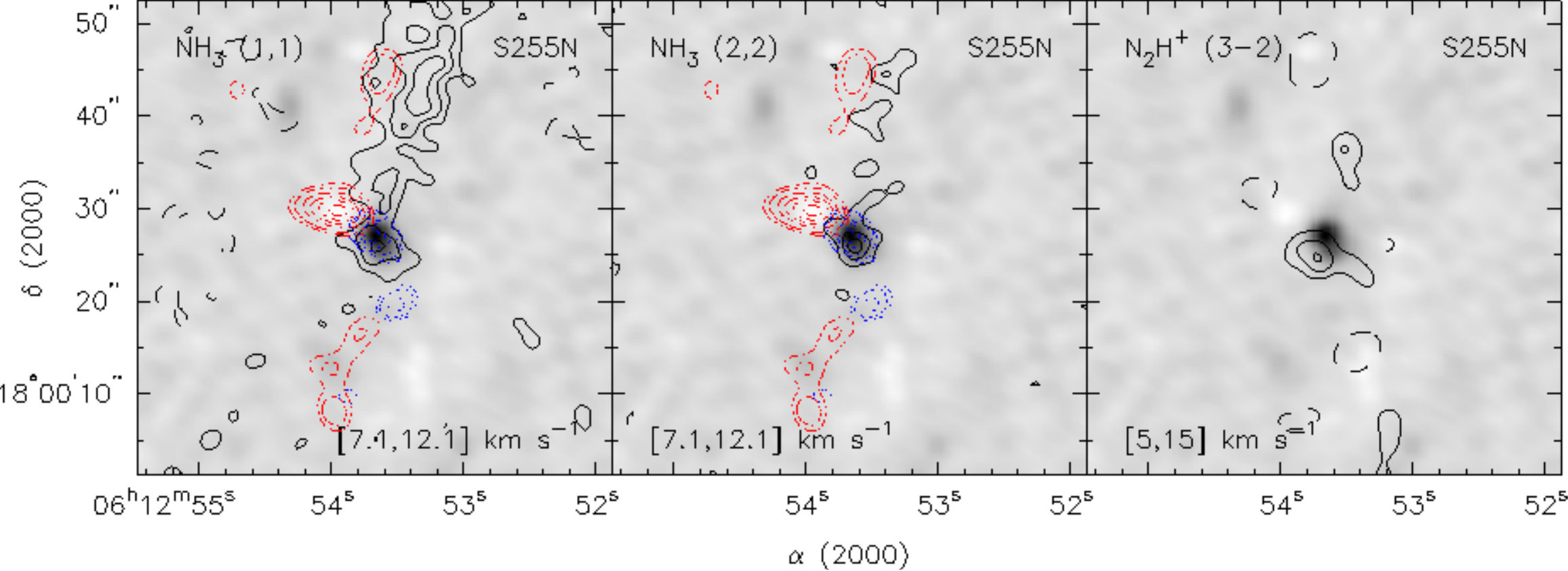} 
\caption{Maps of the NH$_3$ (1,1), NH$_3$ (2,2) and N$_2$H$^+$ (3--2) integrated line emission in the S255N area (contours) overlaid on the 1.1~mm continuum image. The velocity ranges are indicated at the bottom of the plots. The contour levels are (--3, 3, 5, 7)$\times 10$~mJy\,beam$^{-1}$\,km\,s$^{-1}$ for NH$_3$ and (--3, 3, 5, 7)$\times 1.4$~Jy\,beam$^{-1}$\,km\,s$^{-1}$ for N$_2$H$^+$. The dashed contours show negative features due to the missing flux. The dash-dotted (red in the color version) and dotted (blue in the color version) contours show the red-shifted and blue-shifted high-velocity CO emission, respectively.}\label{fig:n_fig1} 
\end{figure*}

\begin{figure}
\includegraphics[width=\columnwidth]{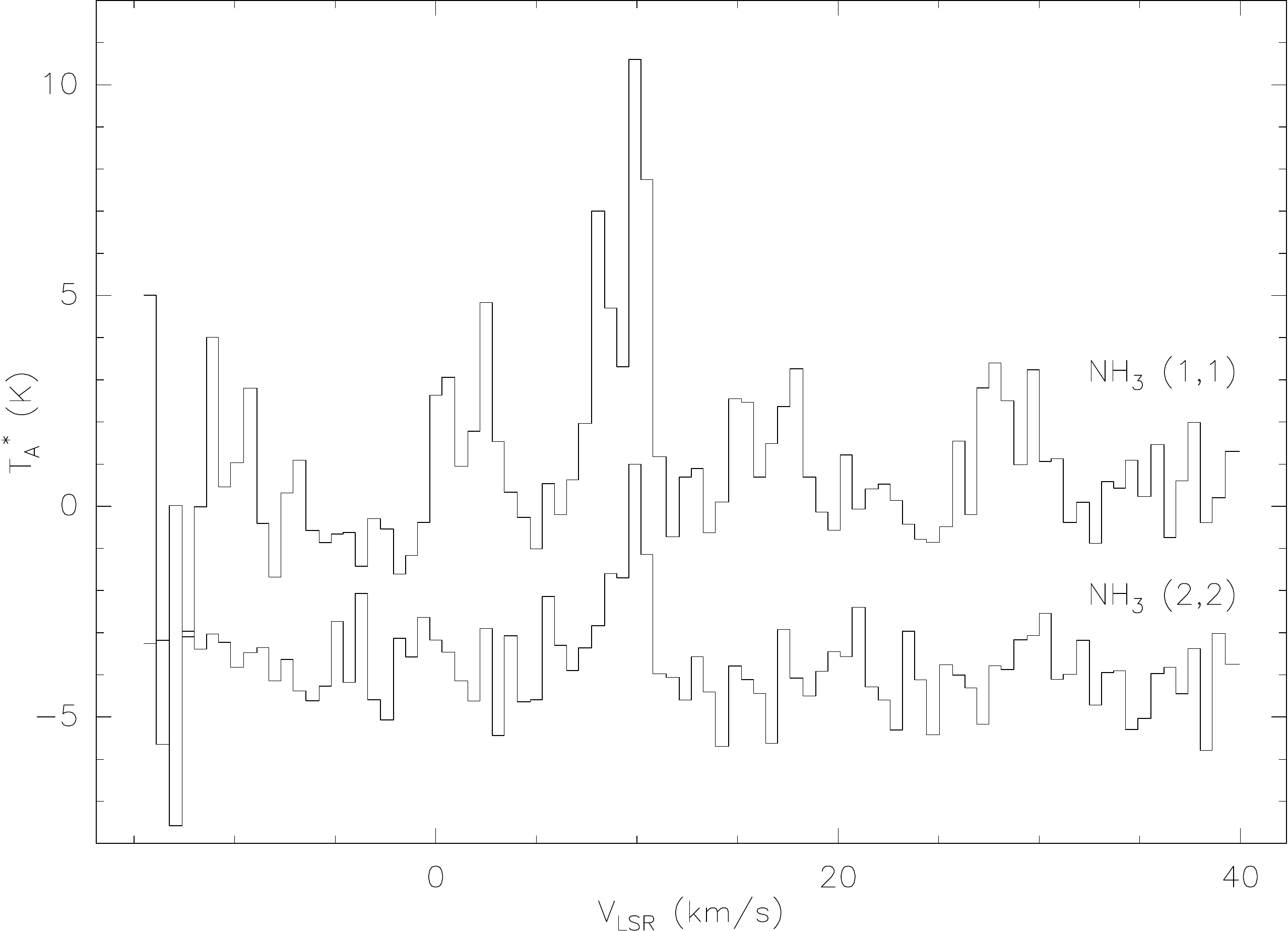} 
\caption{The spectra of the NH$_3$ (1,1) and (2,2) emission from the S255N-NH$_3$ area averaged over the region $2\farcs5\times 3\farcs5$ centered at R.A.(2000) = 06$^h$12$^m$53$^s$.39, Dec.(2000) = 18$^\circ$00$'$41$''$.0 (in units of brightness temperature --- the conversion factor is approximately 320~K per Jy~beam$^{-1}$). The (2,2) spectrum is shifted by --4~K along the vertical axis for clarity.}\label{fig:n_nh3} 
\end{figure}

The ammonia emission is relatively strong in S255N. In both the (1,1) and (2,2) transitions it extends far to the north from the main continuum clump (Fig.~\ref{fig:n_fig1}). This extension has no counterpart in the continuum (at any observed wavelength) and almost no counterpart in other molecular lines (only N$_2$H$^+$ and perhaps DCO$^+$ show similar features, although less extended --- see below). The (1,1) and (2,2) distributions in this northern extension are somewhat different; we discuss this in Section~\ref{sec:disc}. We shall designate this northern source of ammonia emission as S255N-NH$_3$. The (1,1) and (2,2) spectra from this area are presented in Fig.~\ref{fig:n_nh3}. They clearly show two narrow (linewidths of about 1~km\,s$^{-1}$) velocity components (at approximately 10.1 and 8.2~km\,s$^{-1}$). The two velocity components have significantly different line ratios [$ I(2,2)/I(1,1)\sim 0.2$ and $ \sim 0.4 $ for the 8 and 10~km\,s$^{-1}$ components, respectively], implying different temperatures for these two kinematic components. The temperature estimates are given in Section~\ref{sec:disc}.

Another peak of ammonia emission --- the strongest one in the (2,2) transition --- is displaced by a few arc seconds to the south from the main continuum source. Only the 10~km\,s$^{-1}$ kinematic component is readily seen here, although there is a hint on the presence of the 8~km\,s$^{-1}$ component in the (2,2) line (at about 1.5$\sigma$ level). If real, it would imply a significant increase of the temperature of this component in this area since the $ I(2,2)/I(1,1) $ ratio appears to be $ \ga 1 $.

\subsubsection{N$_2$H$^+$}
\begin{figure*}
\includegraphics[width=\textwidth]{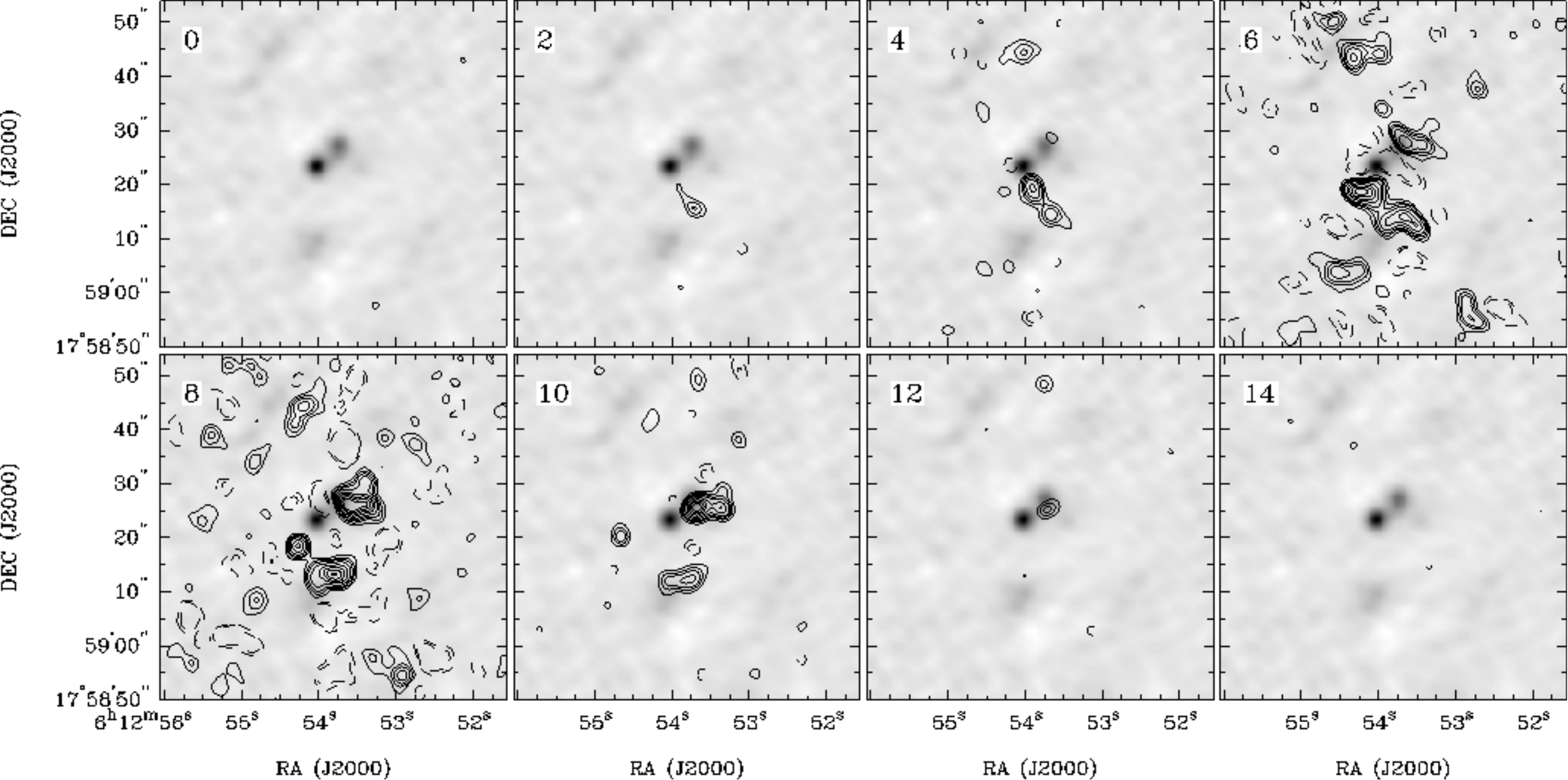} 
\caption{Channel maps of the N$_2$H$^+$ (3--2) line emission in the S255IR area (contours) overlaid on the 1.1~mm continuum image. The numbers in the upper left corner indicate the channel velocity in km\,s$^{-1}$. The contour levels are (--5, --3, 3, 5, 7, 10, 15, 20)$\times 100$~mJy\,beam$^{-1}$. The dashed contours show negative features due to the missing flux. The maps at the central velocities (6--10~km\,s$^{-1}$) may show spurious features due to a limited dynamical range of the observations.}\label{fig:ir_n2h-chmap} 
\end{figure*}

\begin{figure*}
\includegraphics[width=\textwidth]{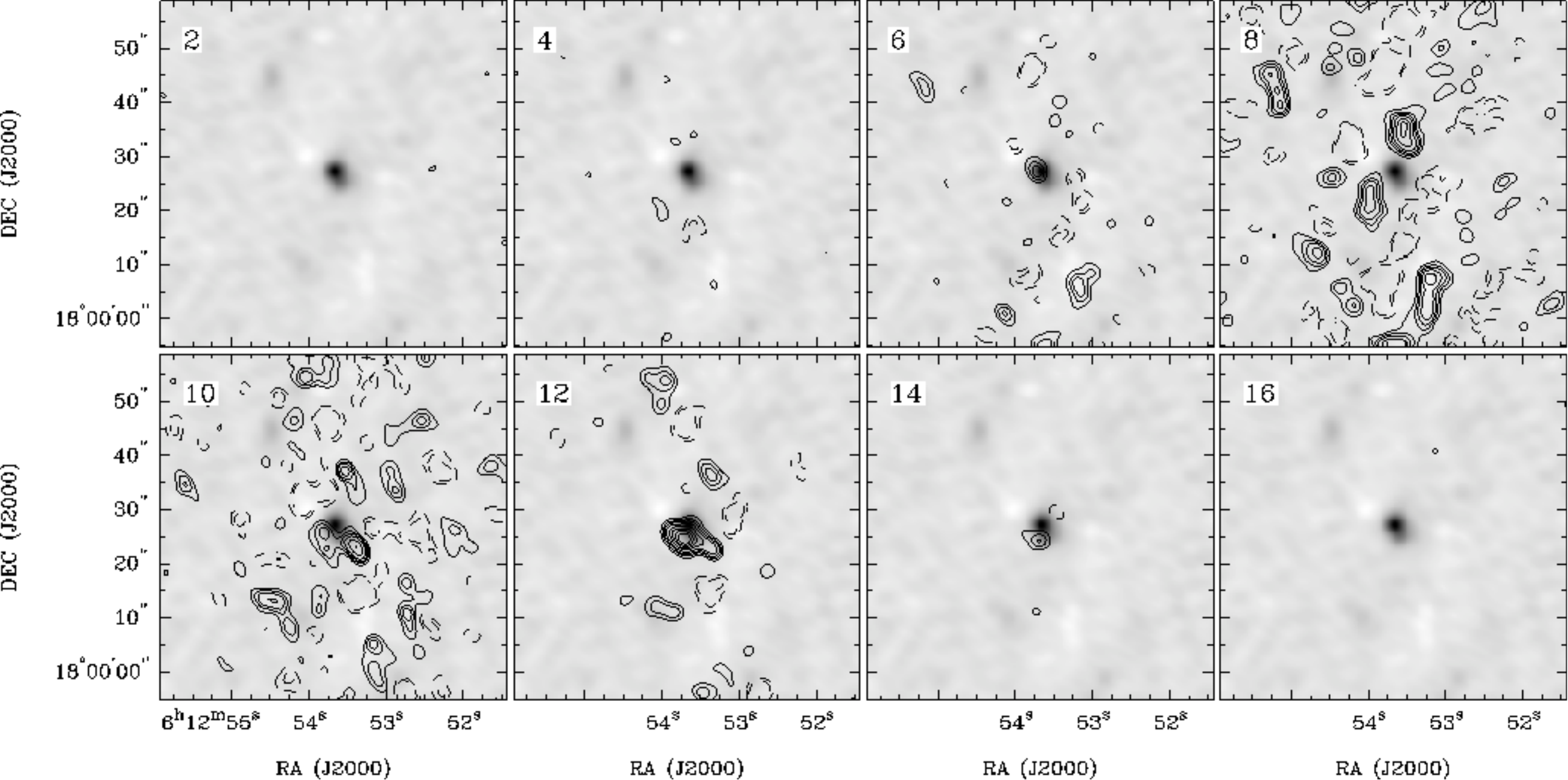} 
\caption{Channel maps of the N$_2$H$^+$ (3--2) line emission in the S255N area (contours) overlaid on the 1.1~mm continuum image. The numbers in the upper left corner indicate the channel velocity in km\,s$^{-1}$. The contour levels are (--5, --3, 3, 5, 7, 10, 15, 20)$\times 100$~mJy\,beam$^{-1}$. The dashed contours show negative features due to the missing flux. The maps at the central velocities (6--10~km\,s$^{-1}$) may show spurious features due to a limited dynamical range of the observations.}\label{fig:n_n2h-chmap} 
\end{figure*}

N$_2$H$^+$ $ J=3-2 $ emission is quite strong in both S255IR and S255N. The integrated intensity maps are shown in Figs.~\ref{fig:ir_fig1},\ref{fig:n_fig1}. To better characterize the structure and kinematics of these regions, we present in Figs.~\ref{fig:ir_n2h-chmap},\ref{fig:n_n2h-chmap} the channel maps of this line with 2~km\,s$^{-1}$ resolution.

In S255IR, while there is no detectable N$_2$H$^+$ emission directly associated with the SMA1 clump, several N$_2$H$^+$ emission peaks can be seen in this general area  (Fig.~\ref{fig:ir_fig1}). Two of these are related to the SMA2 and SMA4 clumps, although the N$_2$H$^+$ peaks do not coincide exactly with the continuum peaks. Two other peaks have no associated continuum. One of these is seen as an extension to the west from the SMA2 clump and is apparently related to the CH$_3$OH emission feature noticed by \citet{Wang11}. This feature is seen in several other methanol lines in our data. NH$_3$ (1,1) emission is also observed here, as are some other lines (e.g., H$_2$CO). We shall designate this source as S255IR-N$_2$H$^+$(1). Another quite strong N$_2$H$^+$ peak, shifted to the SE from SMA1, has no counterpart in the continuum or in any observed lines. For this emission feature we shall use the name S255IR-N$_2$H$^+$(2). Both features are marked in Fig.~\ref{fig:ir_fig1}.

\begin{figure*}
\begin{minipage}{0.32\textwidth}
\centering
\includegraphics[width=\textwidth]{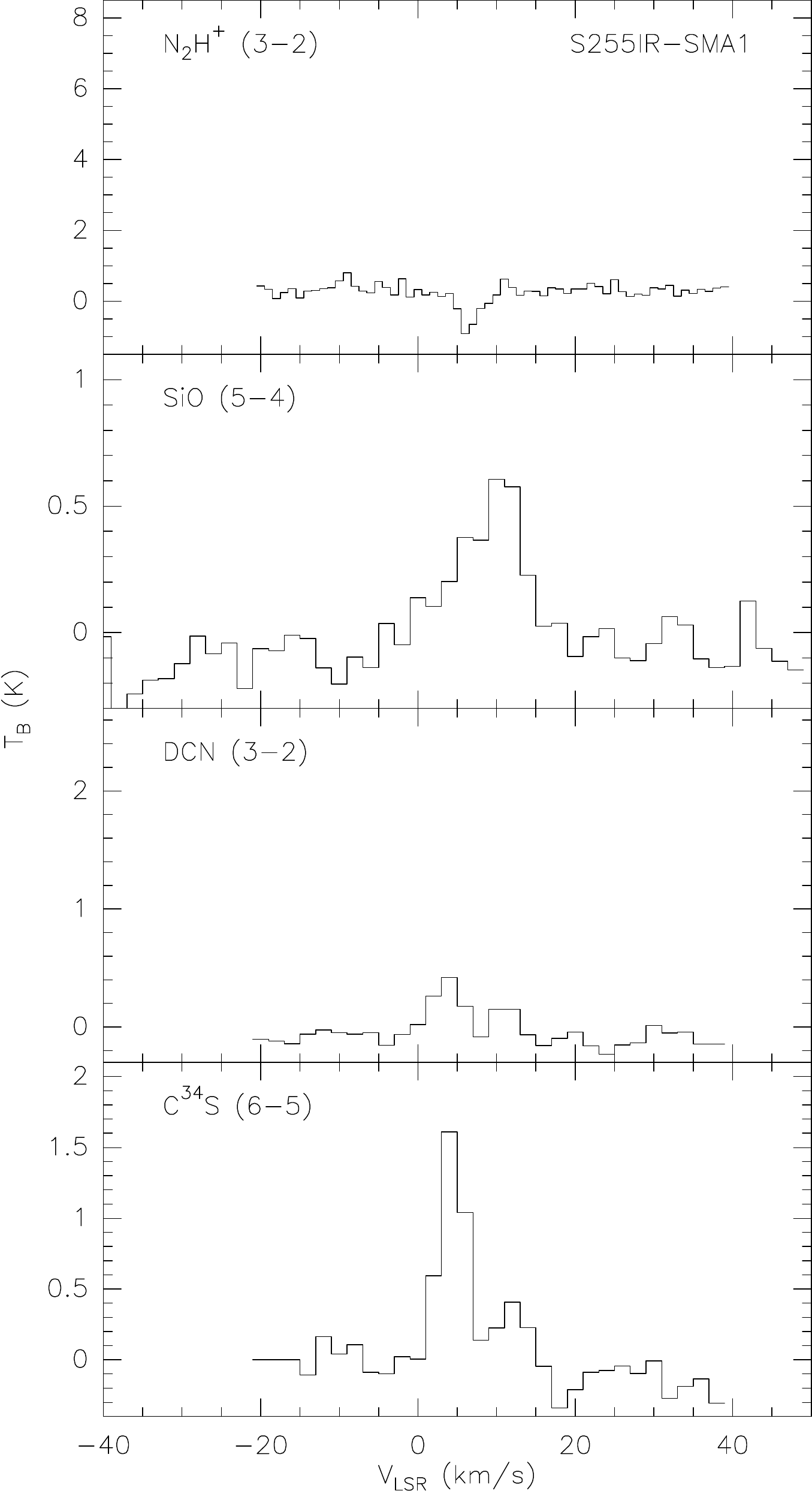}
\end{minipage}
\hfill
\begin{minipage}{0.32\textwidth}
\centering
\includegraphics[width=\textwidth]{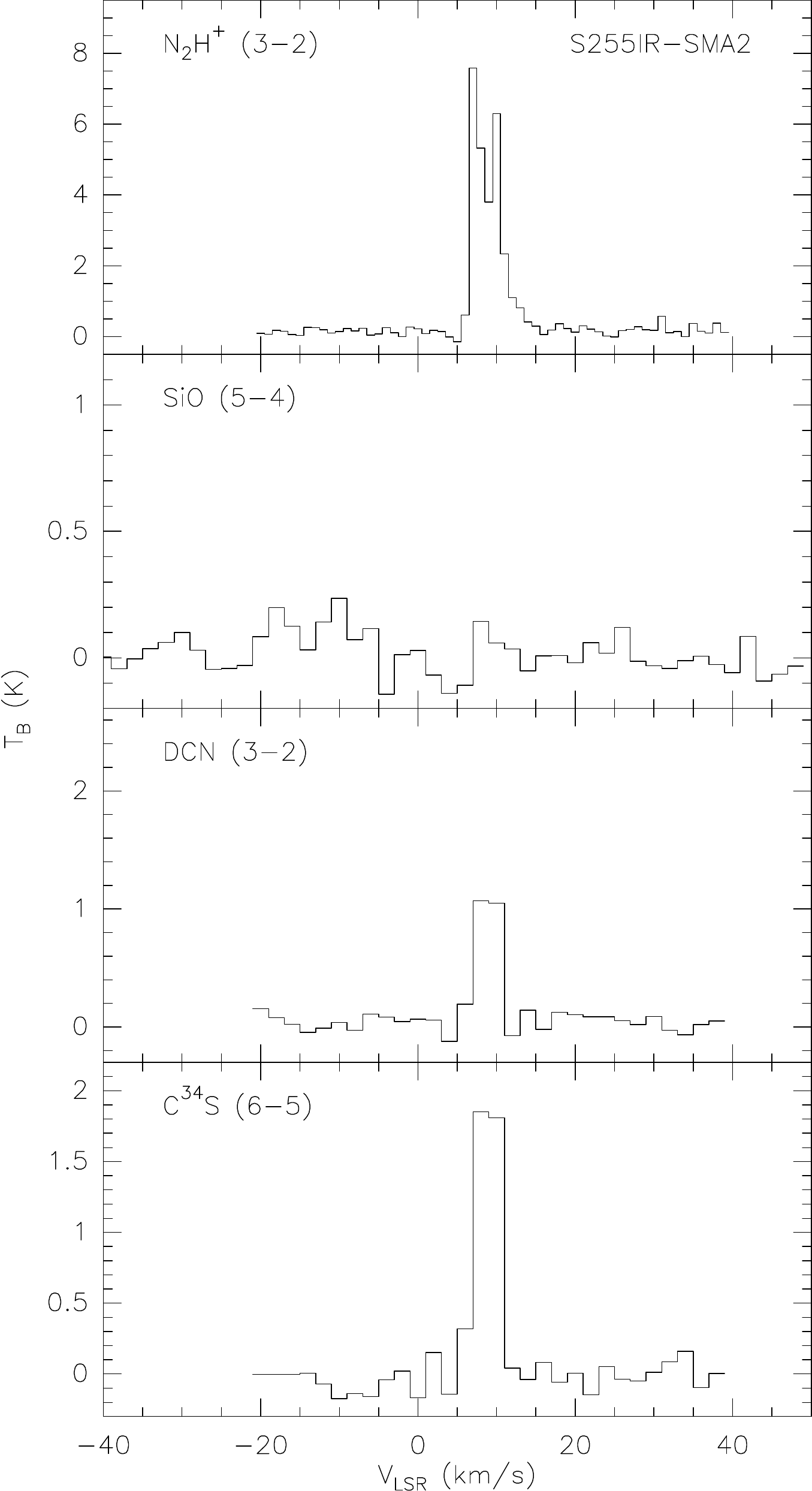}
\end{minipage}
\hfill
\begin{minipage}{0.32\textwidth}
\centering
\includegraphics[width=\textwidth]{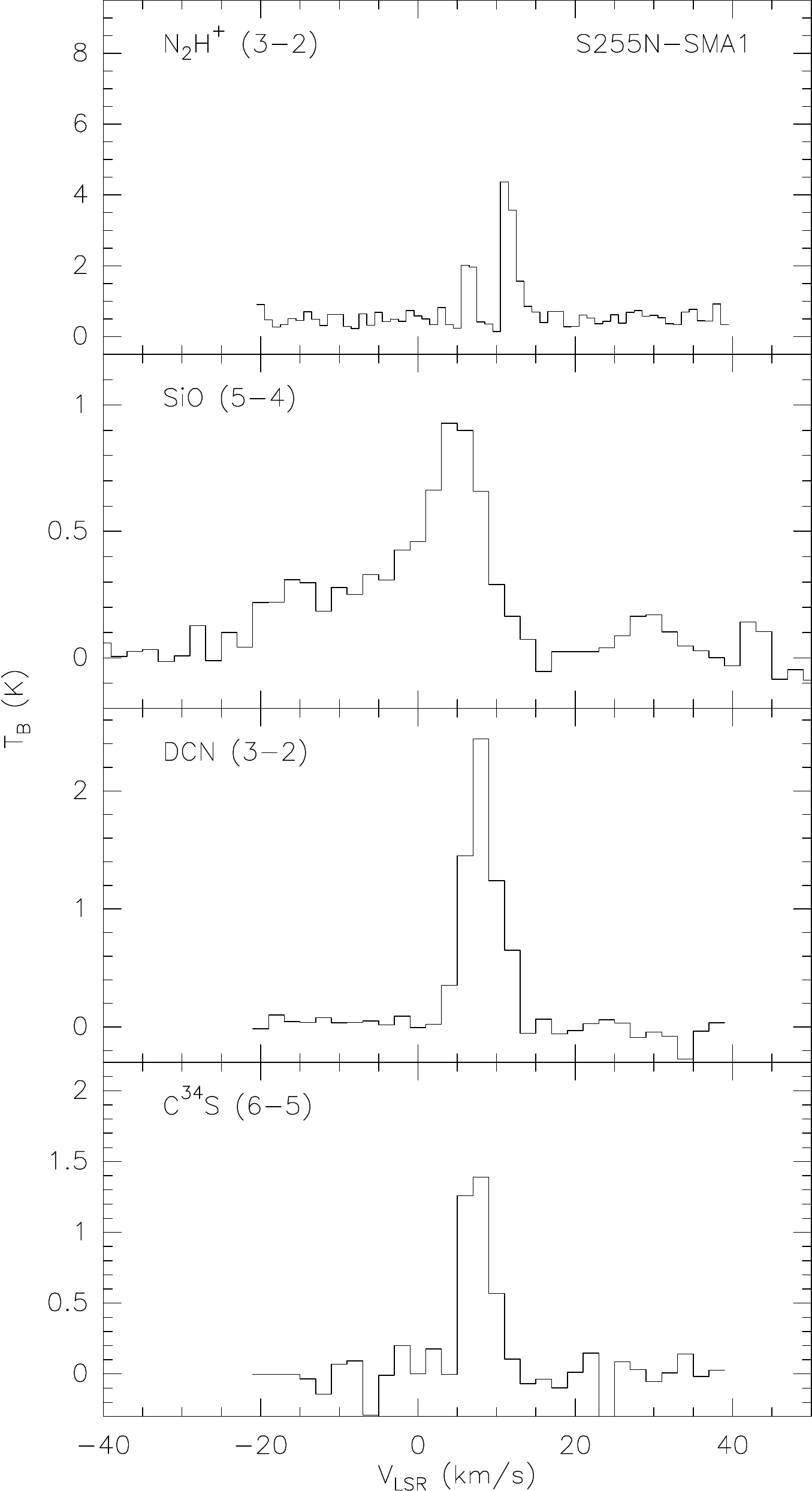}
\end{minipage}
\caption{The spectra of the N$_2$H$^+$ $ J=3-2 $, SiO $ J=5-4 $, DCN $ J=3-2 $, and C$^{34}$S $ J=6-5 $ emission averaged over the regions $2\farcs5\times 2\farcs5$ centered at S255IR-SMA1, S255IR-SMA2 and S255N-SMA1 (in units of brightness temperature).}
\label{fig:spectra}
\end{figure*}

In the S255N area the N$_2$H$^+$ emission shows an extension in the same direction as NH$_3$ (Fig.~\ref{fig:n_fig1}). The main peak of the N$_2$H$^+$ emission is displaced from the continuum peak (similar to the ammonia) and has an extension roughly in the direction of the outflow described by \citet{Wang11}. There is one more N$_2$H$^+$ emission peak about 25$^{\prime\prime}$ to the SSW from S255N-SMA1; it is associated with weak continuum emission (S255N-SMA6) and is traced by several other molecular lines (see below). The spectra of N$_2$H$^+$ and other molecules from S255N-SMA1 are shown in Fig.~\ref{fig:spectra}.


\subsubsection{DCN, DNC and DCO$^+$} \label{sec:dcn}
\begin{figure*}
\includegraphics[width=\textwidth]{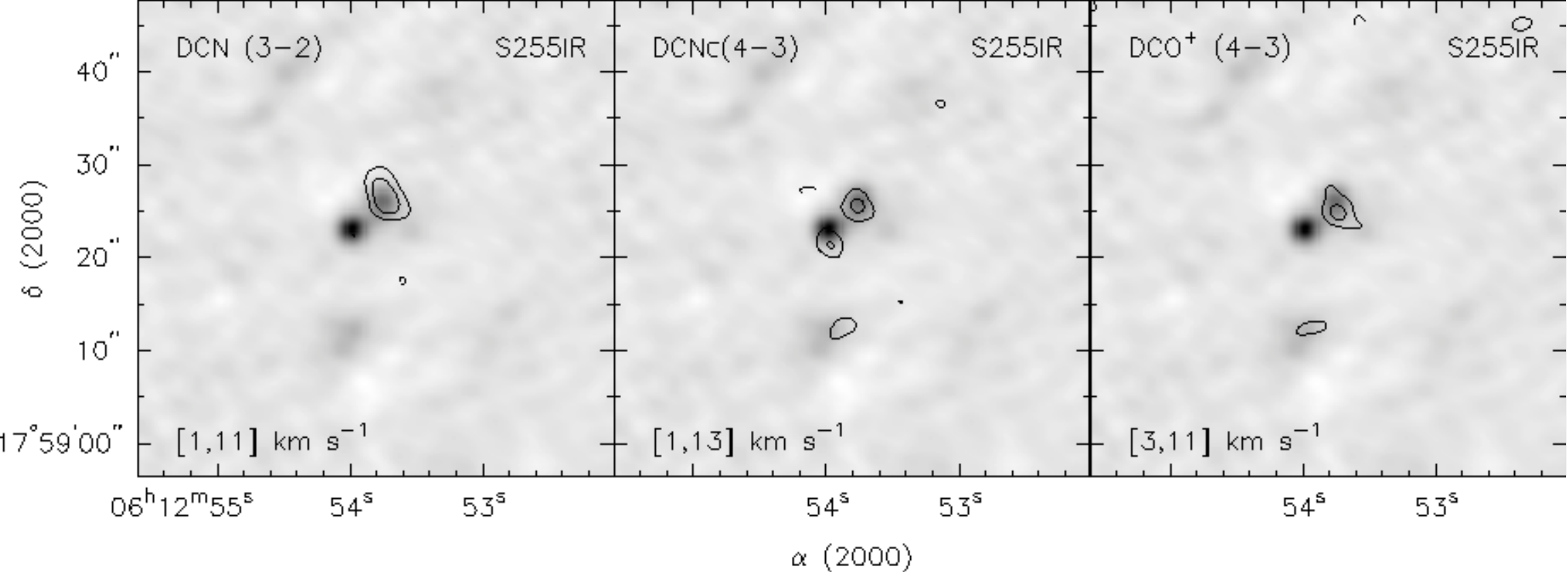} 
\caption{Maps of the DCN $ J=3-2 $ and $ J=4-3 $, and DCO$^+$ $ J=4-3 $ integrated line emission in the S255IR area (contours) overlaid on the 1.1~mm continuum image. The velocity ranges are indicated at the bottom of the plots. The contour levels are (--3, 3, 5)$\times 0.4$~Jy\,beam$^{-1}$\,km\,s$^{-1}$ for DCN $ J=3-2 $ and DCO$^+$ $ J=4-3 $, and  (--3, 3, 5)$\times 0.55$~Jy\,beam$^{-1}$\,km\,s$^{-1}$ for DCN $ J=4-3 $. The dashed contours show negative features due to the missing flux.}\label{fig:ir_fig2} 
\end{figure*}

\begin{figure*}
\includegraphics[width=\textwidth]{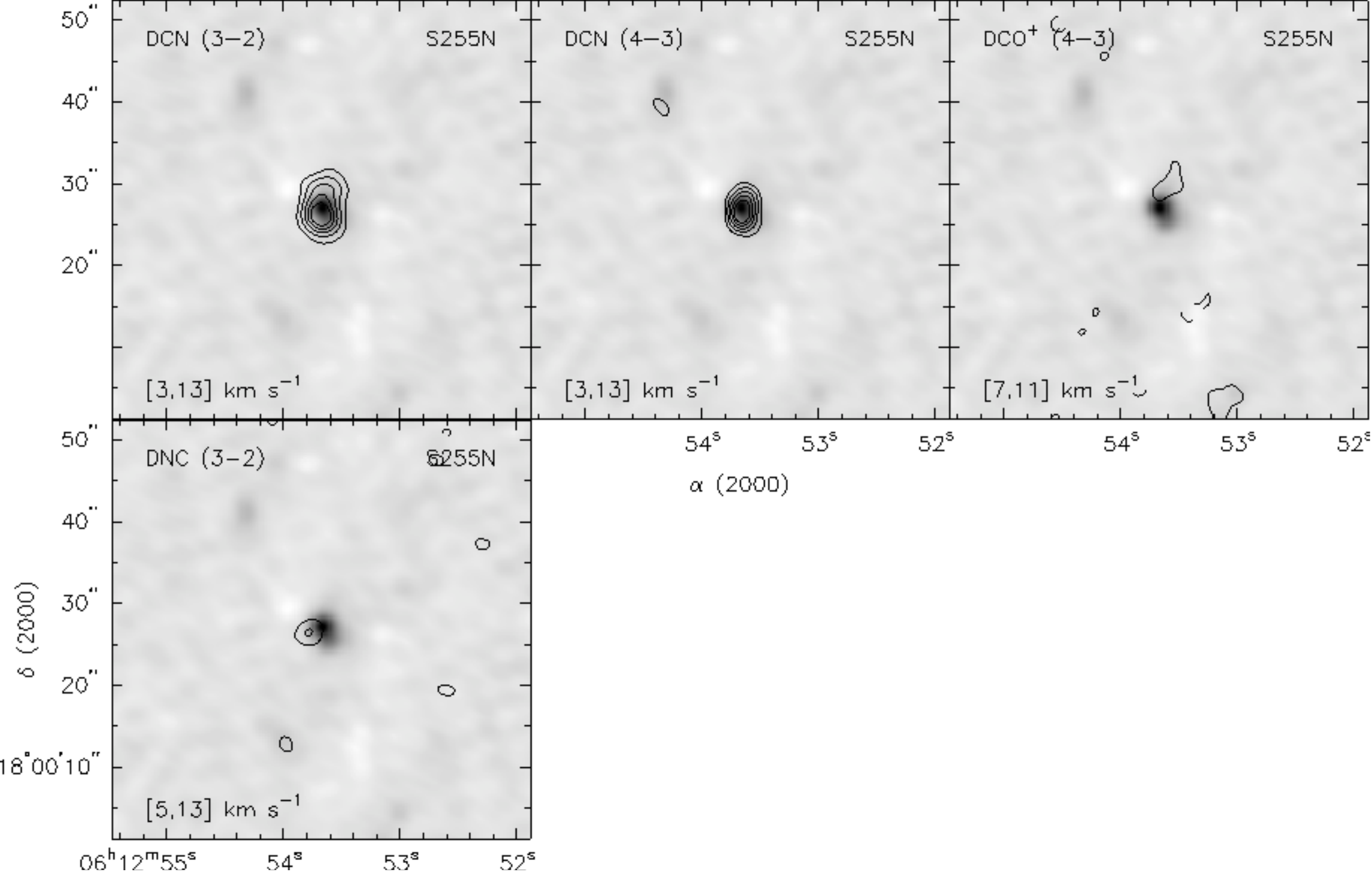} 
\caption{Maps of the DCN $ J=3-2 $ and $ J=4-3 $, DNC $ J=3-2 $ and DCO$^+$ $ J=4-3 $ integrated line emission in the S255N area (contours) overlaid on the 1.1~mm continuum image. The velocity ranges are indicated at the bottom of the plots. The contour levels are (--3, 3, 5, 7, 9, 11)$\times 0.5$~Jy\,beam$^{-1}$\,km\,s$^{-1}$ for DCN $ J=3-2 $, (--3, 3, 5, 7, 9, 11)$\times 0.6$~Jy\,beam$^{-1}$\,km\,s$^{-1}$ for DCN $ J=4-3 $, and (--3, 3, 5)$\times 0.25$~Jy\,beam$^{-1}$\,km\,s$^{-1}$ for DNC $ J=3-2 $ and DCO$^+$ $ J=4-3 $. The dashed contours show negative features due to the missing flux.}\label{fig:n_fig2} 
\end{figure*}

Our data set contains lines of several deuterated molecules: DCN $ J=3-2 $ and $ J=4-3 $, DNC $ J=3-2$, and DCO$^+$ $ J=4-3 $. DCN emission is quite strong in both S255IR and S255N. In S255IR it is observed in both main clumps, SMA1 and SMA2, although its peak is shifted from the continuum peak in SMA1 (Fig.~\ref{fig:ir_fig2}). The DCN $J=4-3$ line is also seen in SMA4. In S255N the DCN emission coincides with the SMA1 continuum clump, having an extension to the north (Fig.~\ref{fig:n_fig2}).

The DNC emission is significantly weaker than DCN. We did not detect DNC in the S255IR area. In S255N it is detected, but the peak is shifted from the DCN and continuum peaks. 

DCO$^+$ emission is seen in both areas. Its distribution is significantly different from that of DCN, and qualitatively resembles that of N$_2$H$^+$. In particular, both species avoid ionized regions, and both are observed towards the S255N SMA6 clump (in contrast to most other molecules).

\subsubsection{SiO}
\begin{figure*}
\includegraphics[width=\textwidth]{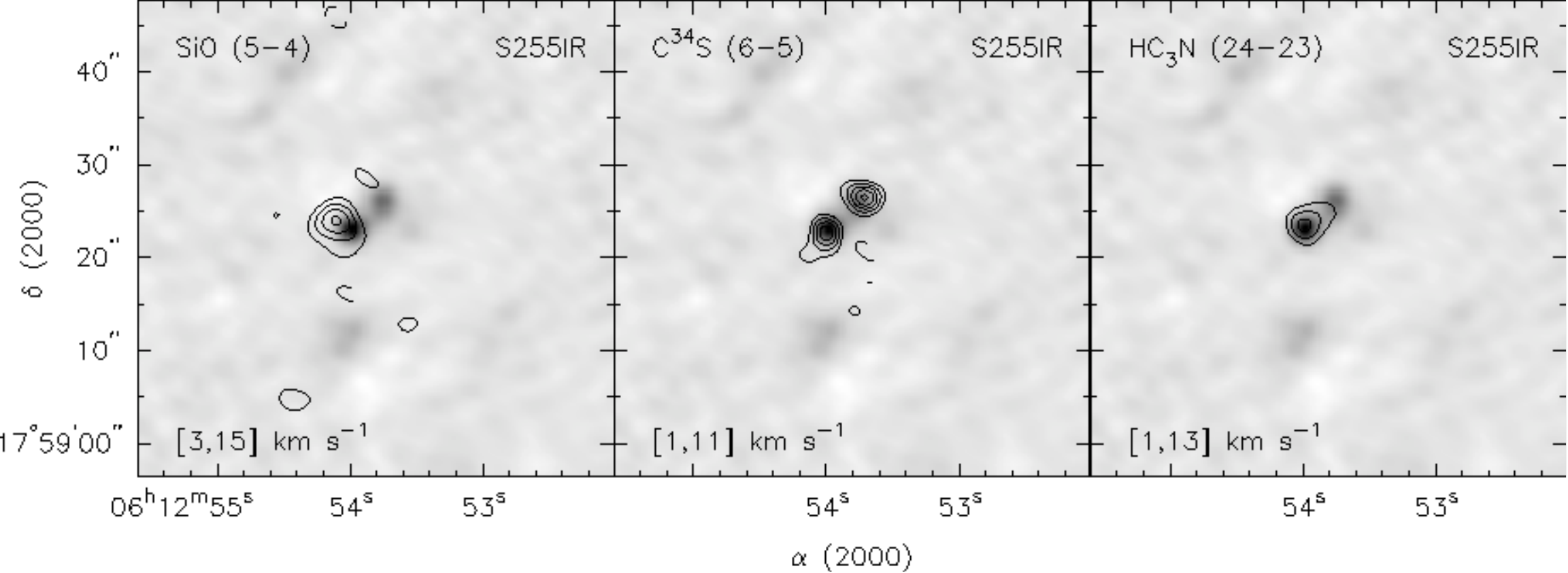} 
\caption{Maps of the SiO $ J=5-4 $, C$^{34}$S $ J=6-5 $ and HC$_3$N $ J=24-23 $ integrated line emission in the S255IR area (contours) overlaid on the 1.1~mm continuum image. The velocity ranges are indicated at the bottom of the plots. The contour levels are (--3, 3, 5, 7, 9)$\times 0.4$~Jy\,beam$^{-1}$\,km\,s$^{-1}$ for SiO $ J=5-4 $ and HC$_3$N $ J=24-23 $, and  (--3, 3, 5, 7)$\times 0.55$~Jy\,beam$^{-1}$\,km\,s$^{-1}$ for C$^{34}$S $ J=6-5 $. The dashed contours show negative features due to the missing flux.}\label{fig:ir_fig3} 
\end{figure*}

As expected, the bulk of the SiO emission in both S255IR and S255N is apparently associated with outflows. In S255IR the strongest SiO emission is observed towards the SMA1 clump (Fig.~\ref{fig:ir_fig3}) which is probably the driving source of the outflow \citep{Wang11}. The peak is shifted from the continuum peak in the direction of the red-shifted outflow lobe. The central velocity is significantly higher than for most of the other lines (Fig.~\ref{fig:spectra}). There is also blue-shifted high-velocity SiO emission. A discussion of the high-velocity emission in various lines is given in Section~\ref{sec:disc}.

S255IR also shows narrow-line SiO emission near SMA2 and SMA4. The possible association of this emission with other features is discussed below. The details of SiO kinematics can be seen in Figs.~\ref{fig:ir_sio-chmap},\ref{fig:n_sio-chmap}. In particular we see SiO emission from the blue-shifted lobe of the extended outflow described by \citet{Wang11}. However, there is no SiO emission from the red-shifted lobe of this outflow.

\begin{figure*}
\includegraphics[width=\textwidth]{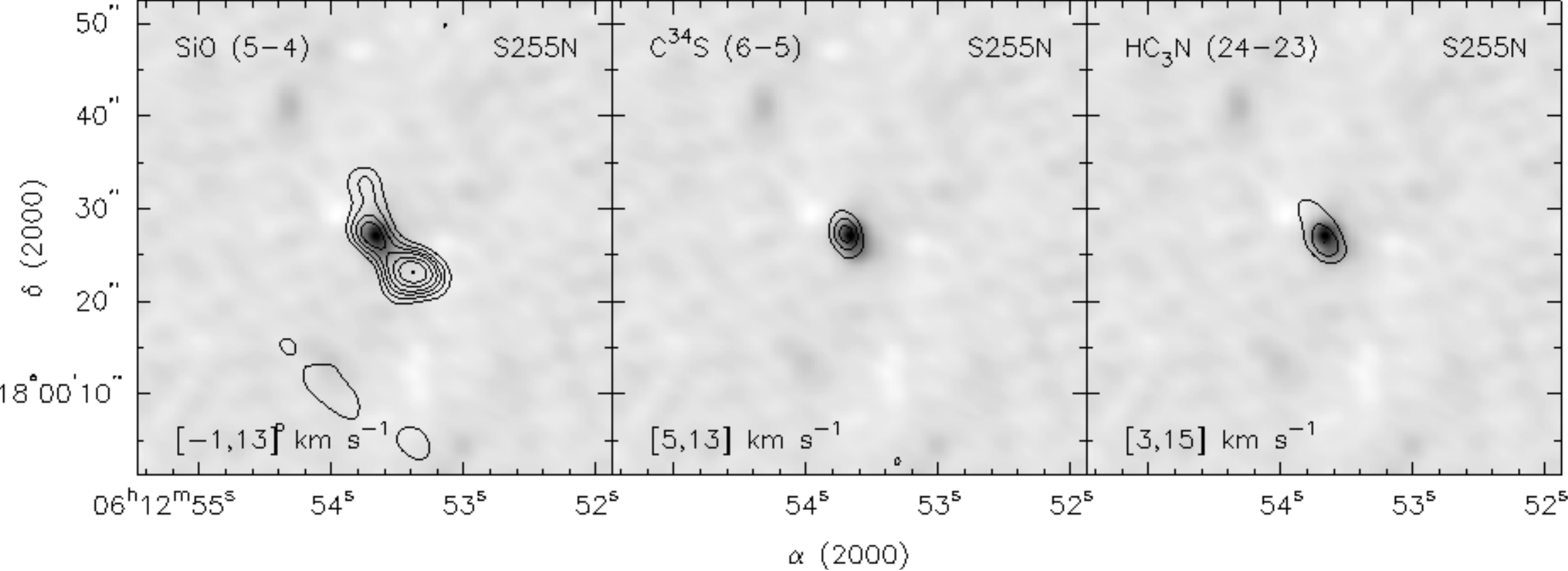} 
\caption{Maps of the SiO $ J=5-4 $, C$^{34}$S $ J=6-5 $ and HC$_3$N $ J=24-23 $ integrated line emission in the S255N area (contours) overlaid on the 1.1~mm continuum image. The velocity ranges are indicated at the bottom of the plots. The contour levels are (--3, 3, 5, 7, 9, 12, 15, 19)$\times 0.4$~Jy\,beam$^{-1}$\,km\,s$^{-1}$ for SiO $ J=5-4 $, (--3, 3, 5, 7)$\times 0.55$~Jy\,beam$^{-1}$\,km\,s$^{-1}$ for C$^{34}$S $ J=6-5 $ and (--3, 3, 5, 7)$\times 0.4$~Jy\,beam$^{-1}$\,km\,s$^{-1}$ for HC$_3$N $ J=24-23 $. The dashed contours show negative features due to the missing flux.}\label{fig:n_fig3} 
\end{figure*}

\begin{figure*}
\includegraphics[width=\textwidth]{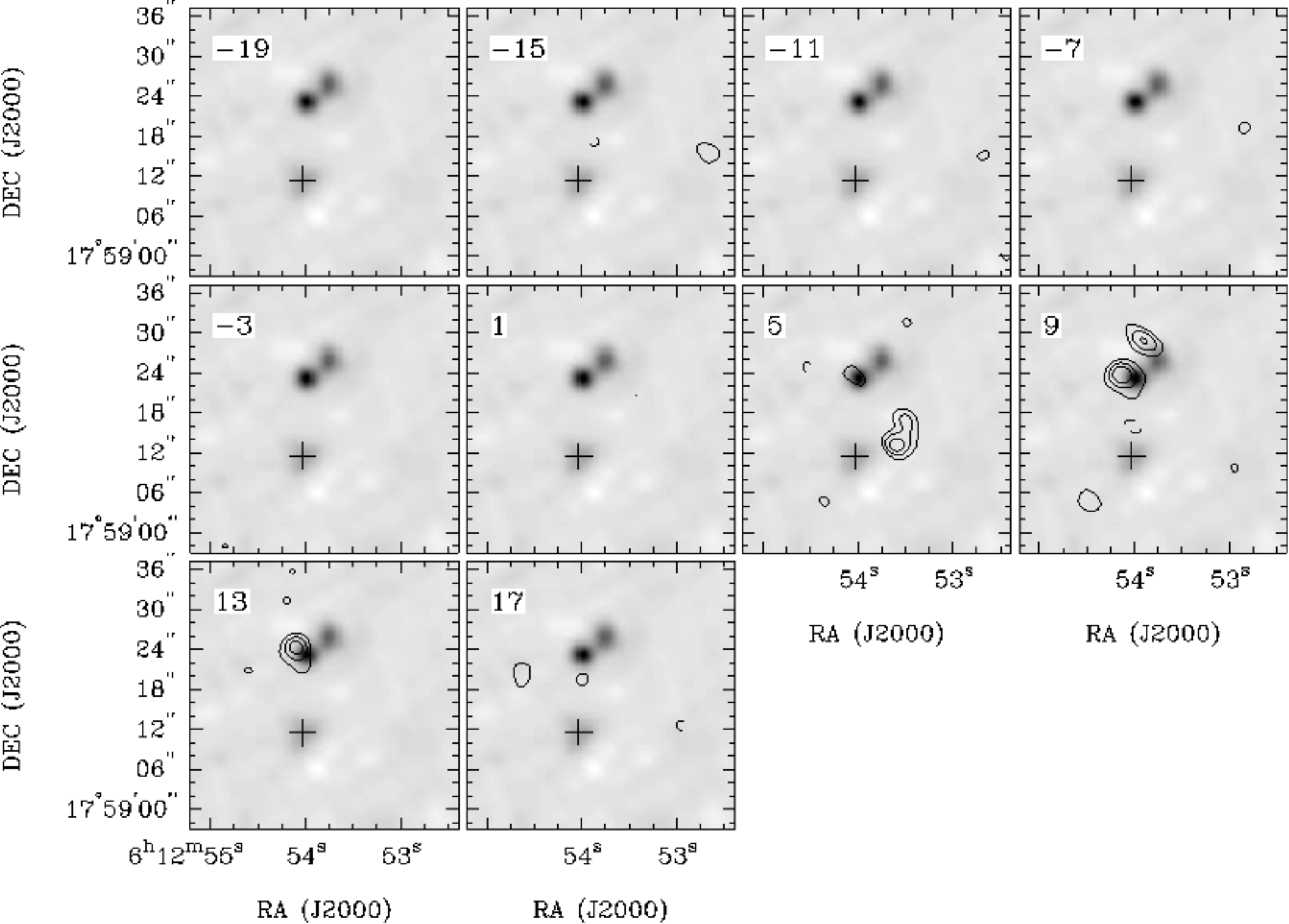} 
\caption{Channel maps of the SiO (5--4) line emission in the S255IR area (contours) overlaid on the 1.1~mm continuum image. The numbers in the upper left corner indicate the channel velocity in km\,s$^{-1}$. The contour levels are (--3, 3, 5, 7, 10, 15, 20)$\times 45$~mJy\,beam$^{-1}$. The dashed contours show negative features due to the missing flux. The cross marks the position of the SMA4 continuum clump.}\label{fig:ir_sio-chmap} 
\end{figure*}

\begin{figure*}
\includegraphics[width=\textwidth]{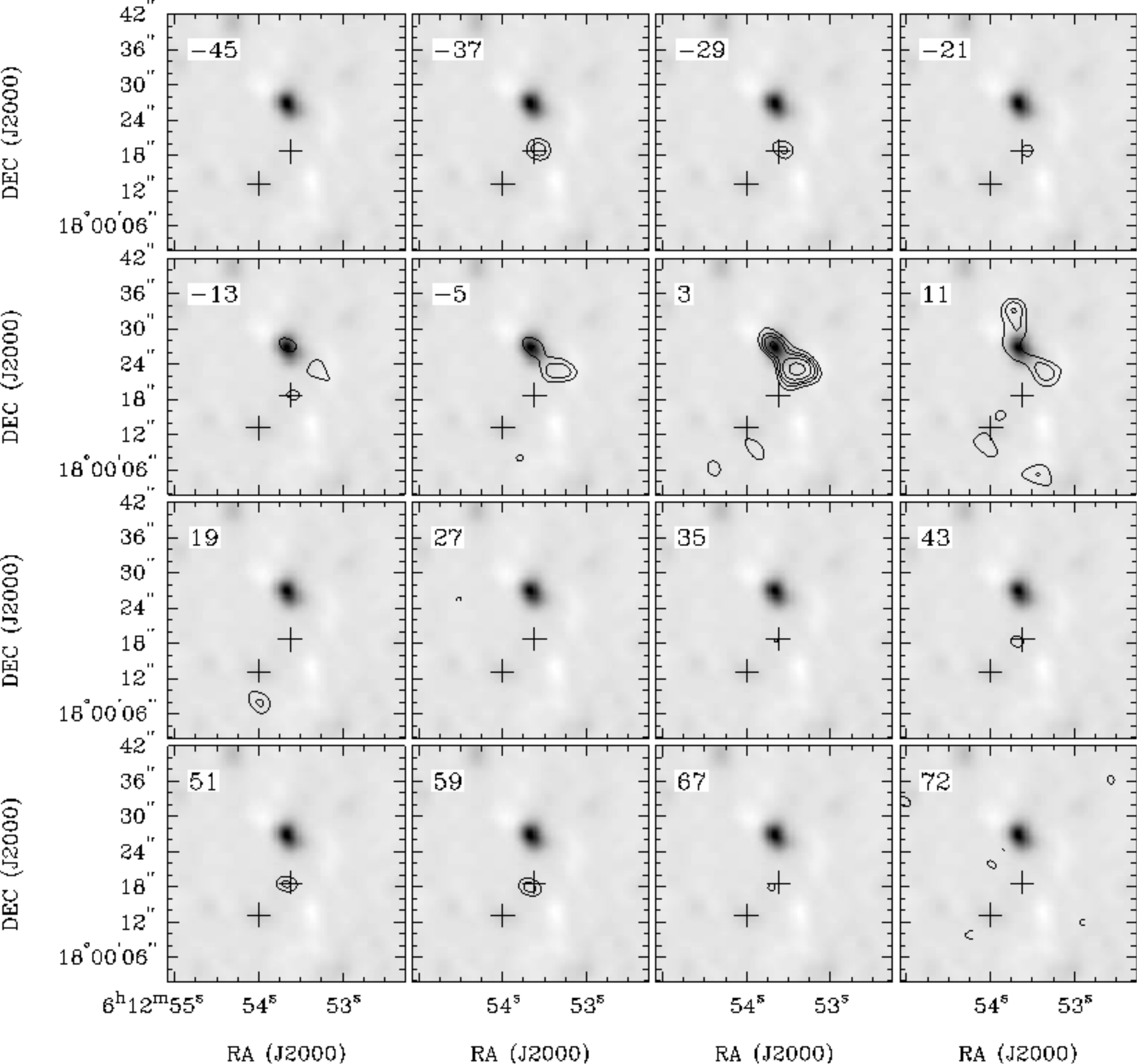} 
\caption{Channel maps of the SiO (5--4) line emission in the S255N area (contours) overlaid on the 1.1~mm continuum image. The numbers in the upper left corner indicate the channel velocity in km\,s$^{-1}$. The contour levels are (--3, 3, 5, 7, 10, 15, 20)$\times 40$~mJy\,beam$^{-1}$. The dashed contours show negative features due to the missing flux. The crosses mark the positions of the SMA3 and SMA5 continuum clumps.}\label{fig:n_sio-chmap} 
\end{figure*}

\begin{figure}
\includegraphics[width=\columnwidth]{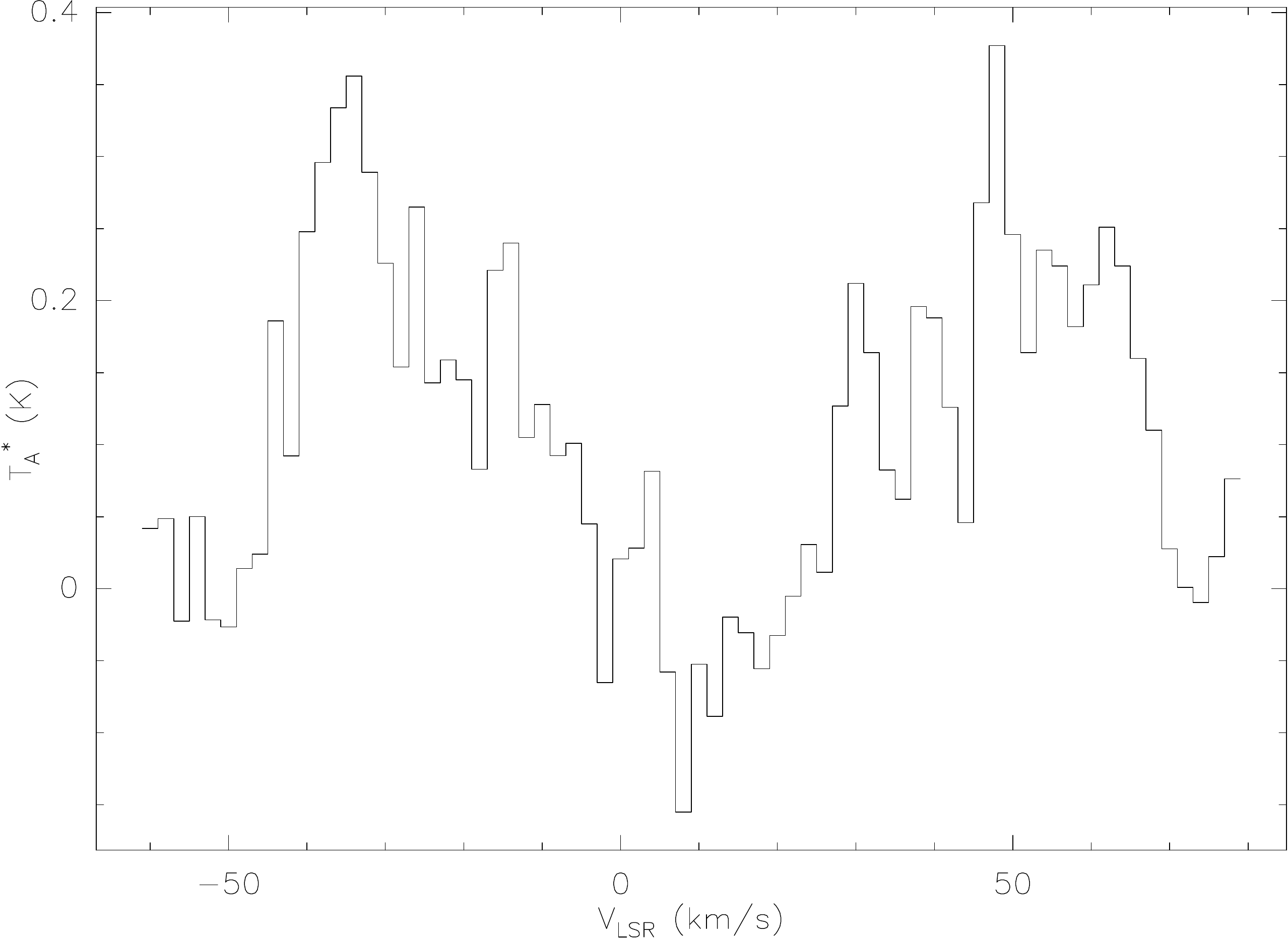} 
\caption{The spectrum of the SiO $ J=5-4 $ emission from the S255N-SMA3 area averaged over the region $3\farcs5\times 2\farcs5$ centered at S255N-SMA3 (in units of brightness temperature --- the conversion factor is approximately 2.2~K per Jy~beam$^{-1}$).}\label{fig:n-mm3_sio} 
\end{figure}

\begin{figure}
\includegraphics[width=\columnwidth]{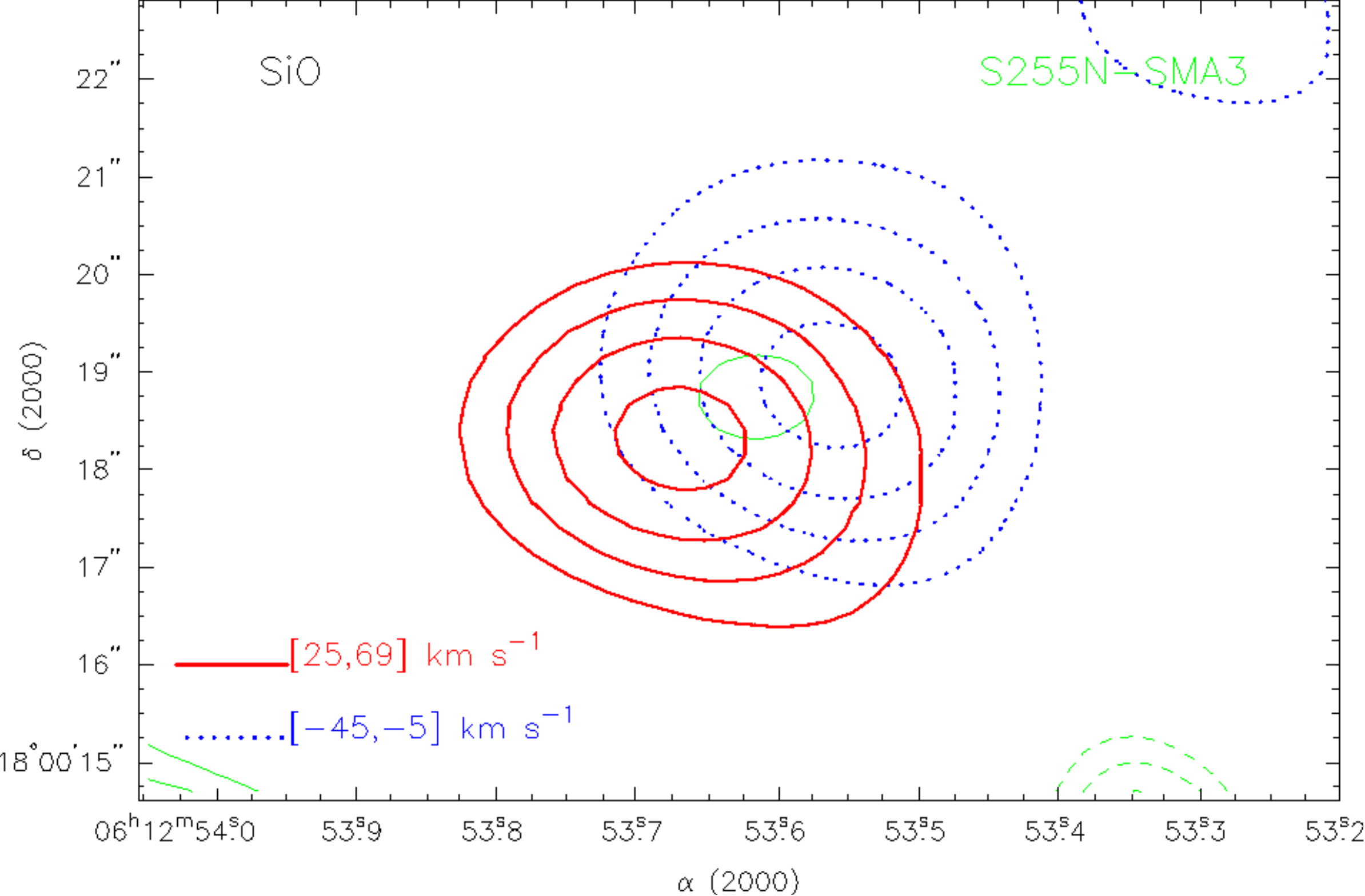} 
\caption{The maps of the blue-shifted (blue thick dotted contours) and red-shifted (red thick solid contours) SiO $ J=5-4 $ emission in the S255N-SMA3 area overlaid with the map of the 1.1~mm continuum emission (green thin contours). The velocity ranges are indicated at the bottom of the plot. The contour levels for SiO are (--3, 3, 5, 7, 9)$\times 0.75$~Jy\,beam$^{-1}$\,km\,s$^{-1}$. The contour parameters for continuum are the same as in Fig.~\ref{fig:NIR}. The dashed contours show negative features due to the missing flux.}\label{fig:n-mm3_sio-map} 
\end{figure}

In S255N (Fig.~\ref{fig:n_fig3}) the main peak of the SiO emission coincides with the SW (blue-shifted) lobe of the molecular outflow traced in CO \citep{Wang11} and other lines. The SiO profile shows an extended blue wing. In addition, there are very broad SiO emission features clearly seen in the SiO channel maps (Fig.~\ref{fig:n_sio-chmap}) associated with the SMA3 and SMA5 clumps. In the vicinity of the SMA3 clump, the SiO emission spans from about --45~km\,s$^{-1}$ to about +65~km\,s$^{-1}$ (Fig.~\ref{fig:n-mm3_sio}). The blue-shifted and red-shifted parts of the emission are spatially separated as can be easily seen in the channel maps (Fig.~\ref{fig:n_sio-chmap}) and more clearly in Fig.~\ref{fig:n-mm3_sio-map}. All this indicates the presence of a high-velocity outflow associated with this clump. This outflow is also traced in high-velocity CO emission (Fig.~\ref{fig:n_fig1} here and Fig.~10 in \citealt{Wang11}).

\begin{figure}
\includegraphics[width=\columnwidth]{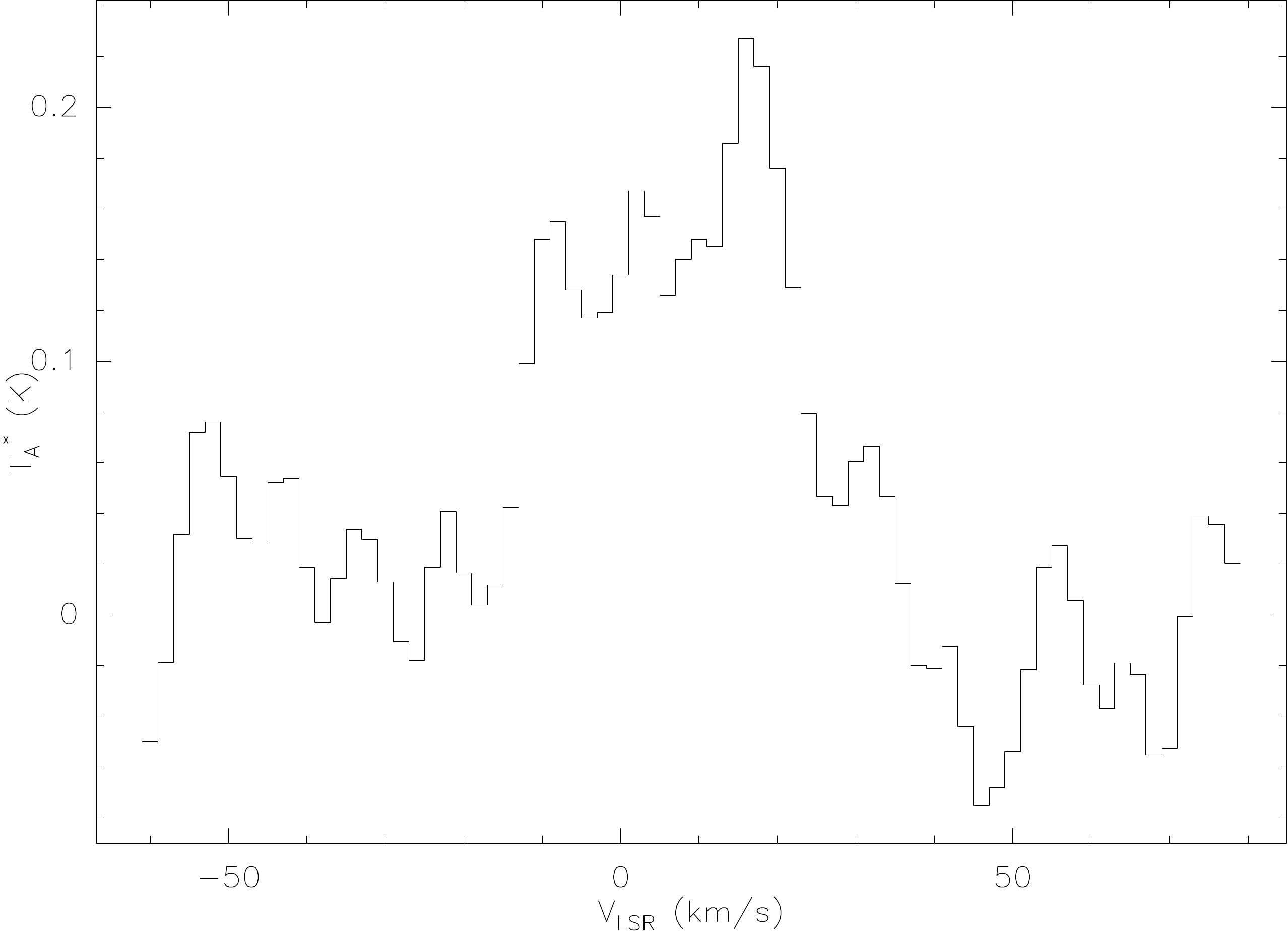} 
\caption{The spectrum of the SiO $ J=5-4 $ emission from the S255N-SMA5 area averaged over the region $5\farcs5\times 5\farcs5$ shifted to the south by $4\farcs7$ from the S255N-SMA5 central position (in units of brightness temperature --- the conversion factor is approximately 2.2~K per Jy~beam$^{-1}$).}\label{fig:n-mm5_sio} 
\end{figure}

\begin{figure}
\includegraphics[width=\columnwidth]{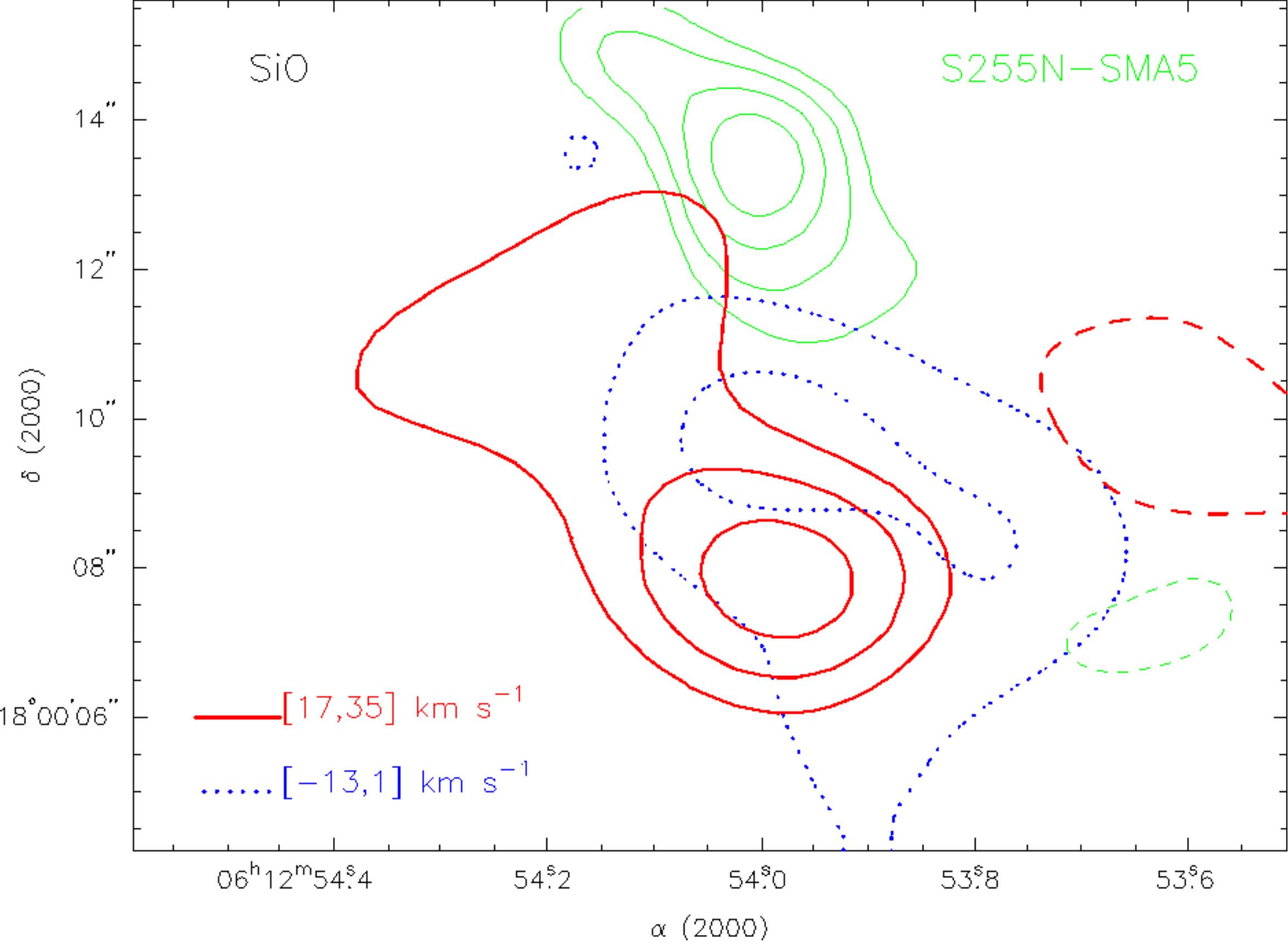} 
\caption{The maps of the blue-shifted (blue thick dotted contours) and red-shifted (red thick solid contours) SiO $ J=5-4 $ emission in the S255N-SMA5 area overlaid with the map of the 1.1~mm continuum emission (green thin contours). The velocity ranges are indicated at the bottom of the plot. The contour levels for SiO are (--3, 3, 5, 7)$\times 0.25$~Jy\,beam$^{-1}$\,km\,s$^{-1}$. The contour parameters for continuum are the same as in Fig.~\ref{fig:NIR}. The dashed contours show negative features due to the missing flux.}\label{fig:n-mm5_sio-map} 
\end{figure}

The total width of the SiO emission associated  with the SMA5 clump is about 50~km\,s$^{-1}$ (Fig.~\ref{fig:n-mm5_sio}). The blue-shifted and red-shifted SiO emissions are spatially separated (Fig.~\ref{fig:n-mm5_sio-map}). This suggests that another outflow is related to this clump. The driving source of the outflow is shifted by a few arcseconds to the south from the SMA5 continuum peak.

\subsubsection{CH$_3$OH}
\begin{figure*}
\includegraphics[width=\textwidth]{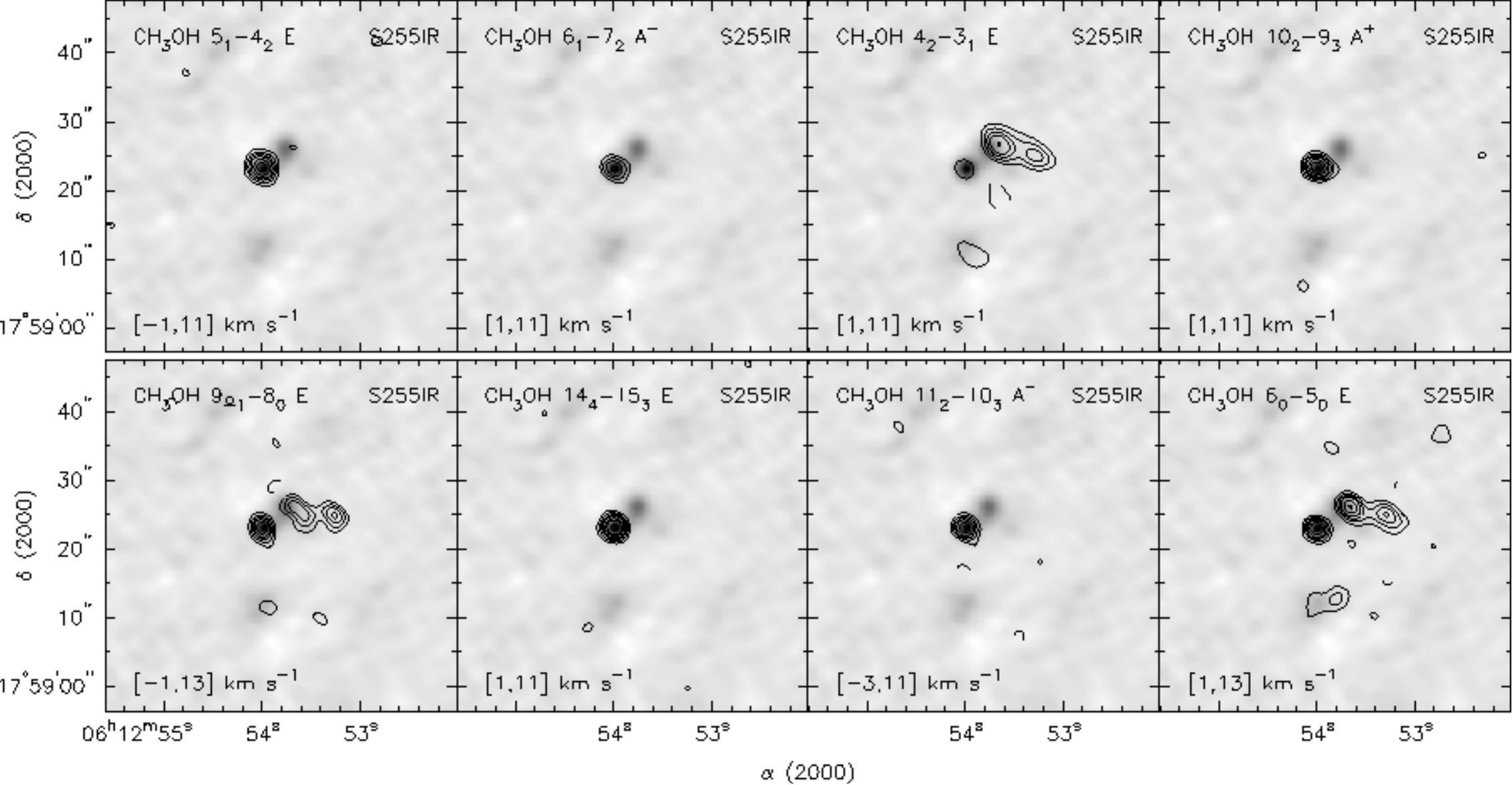} 
\caption{Maps of the CH$_3$OH integrated line emission in the S255IR area (contours) overlaid on the 1.1~mm continuum image. The velocity ranges are indicated at the bottom of the plots. The contour levels are (--3, 3, 5, 7, 9, 12, 15, 20)$\times 0.3$~Jy\,beam$^{-1}$\,km\,s$^{-1}$ for the $ 5_{1}-4_{2} $~E, $ 6_{1}-7_{2} $~A$^-$, $ 10_{2}-9_{3} $~A$^+$ and $ 14_{4}-15_{3} $~E lines, and (--3, 3, 5, 7, 9, 12, 15, 20)$\times 0.6$~Jy\,beam$^{-1}$\,km\,s$^{-1}$ for the other lines. The dashed contours show negative features due to the missing flux.}\label{fig:ir_ch3oh} 
\end{figure*}

\begin{figure*}
\flushleft
\includegraphics[width=\textwidth]{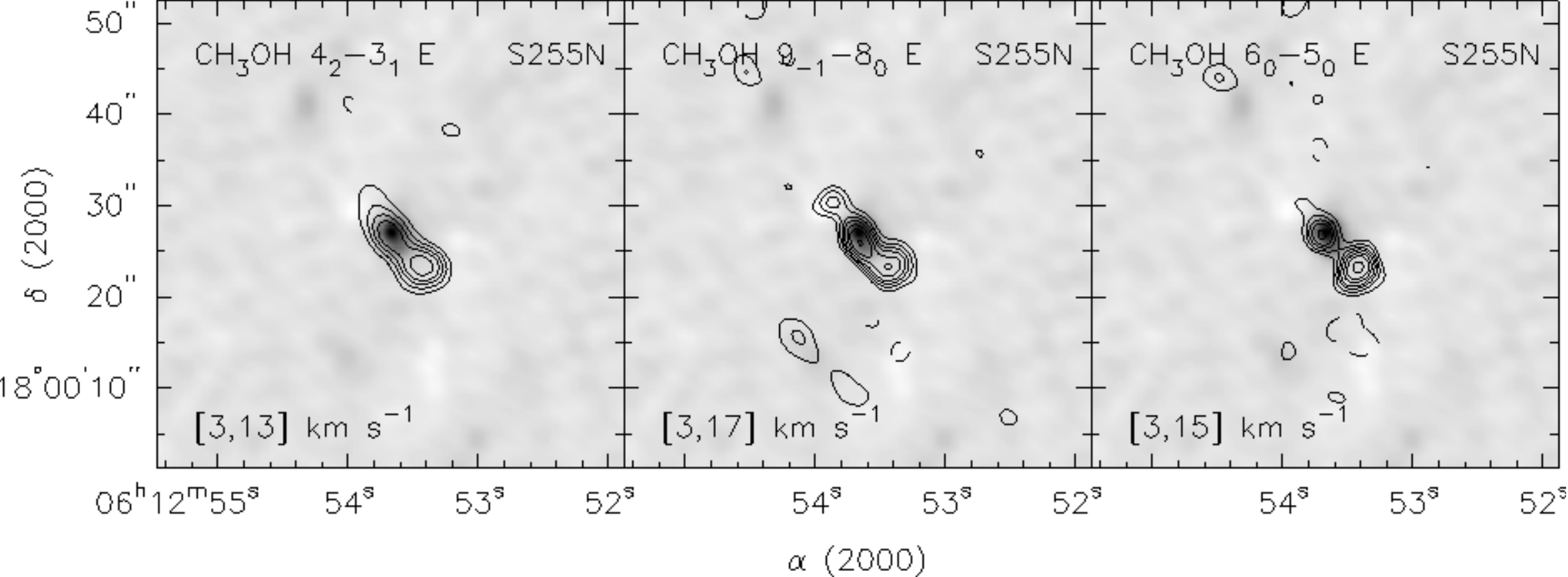} 
\caption{Maps of the CH$_3$OH integrated line emission in the S255N area (contours) overlaid on the 1.1~mm continuum image. The velocity ranges are indicated at the bottom of the plots. The contour levels are (--3, 3, 5, 7, 9, 12, 15, 20)$\times 0.6$~Jy\,beam$^{-1}$\,km\,s$^{-1}$.  The dashed contours show negative features due to the missing flux.}\label{fig:n_ch3oh} 
\end{figure*}

In our band we observed several methanol lines of different excitation. Maps of the methanol emission in the lines detected in the S255IR and S255N areas are presented in Figs.~\ref{fig:ir_ch3oh},\ref{fig:n_ch3oh}.

The main features of the methanol emission in the central parts of S255IR and S255N are the same as reported by \citet{Wang11}. The highest excitation lines are detected only in S255IR-SMA1. Several methanol lines trace some of our new continuum clumps, in particular S255IR-SMA4, S255N-SMA4 and S255N-SMA5. \citet{Wang11} reported an emission in one of the methanol lines from the area which we designate here as S255IR-N$_2$H$^+$(1). Our data show an emission here in 3 other methanol lines.

Methanol lines from S255IR-SMA1 and S255N-SMA1 show wings, although less extended than in SiO. We do not see high velocity methanol emission from S255N-SMA3 and SMA5 where we report SiO outflows.

An inspection of the presented maps indicates apparent peculiarities in the
methanol excitation. For example the low excitation $ 5_{1}-4_{2} $~E line is much stronger in S255-SMA1 than in SMA2, in contrast to other lines of comparable excitation. The same can be said about the $ 9_{-1}-8_{0} $~E line in the S255N area (the component shifted by a few arcseconds to north-east from S255N-SMA1). These and some other methanol lines in our band can be masing \citep[][and A.~Sobolev, private communication]{Sobolev93,Cragg05,Voronkov12}. 

\subsubsection{Other lines}
\begin{figure*}
\includegraphics[width=\textwidth]{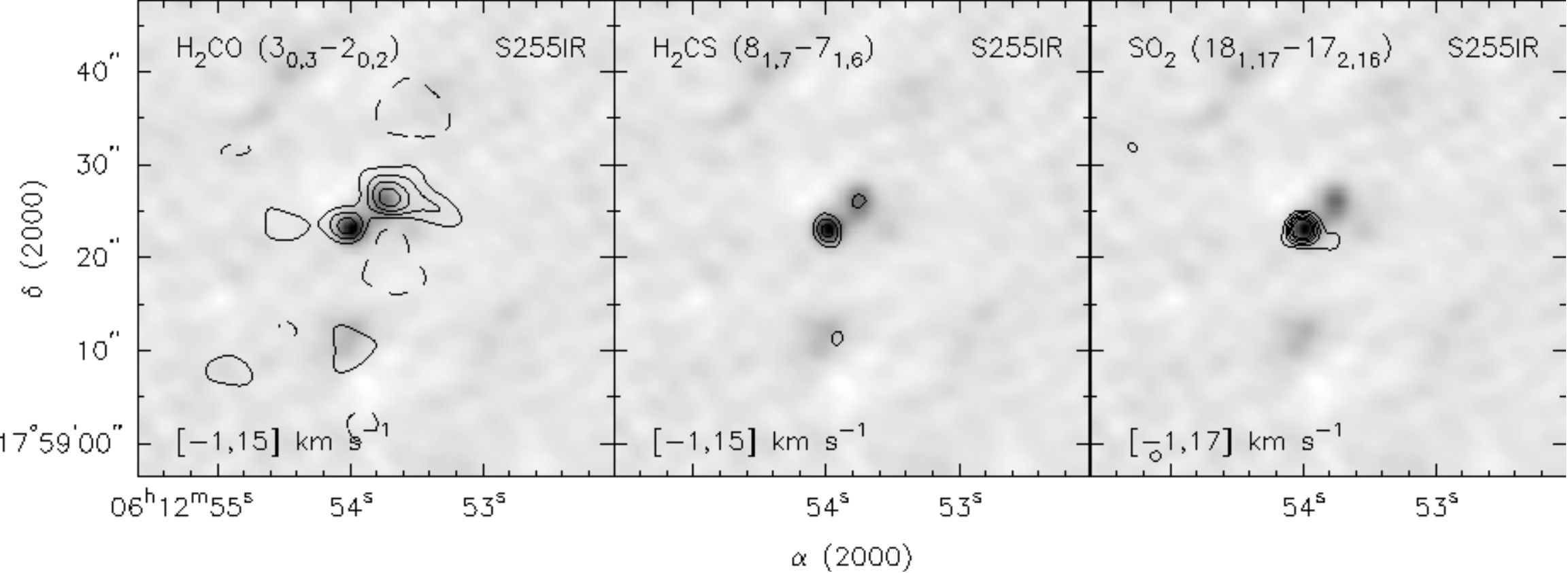} 
\caption{Maps of the H$_2$CO $3_{0,3} -    2_{0,2}$, H$_2$CS $8_{1,7} -    7_{1,6}$ and SO$_2$ 18$_{1,17}-17_{2,16}$ integrated line emission in the S255IR area (contours) overlaid on the 1.1~mm continuum image. The velocity ranges are indicated at the bottom of the plots. The contour levels are (--3, 3, 5, 7, 9)$\times 0.6$~Jy\,beam$^{-1}$\,km\,s$^{-1}$. The dashed contours show negative features due to the missing flux.}\label{fig:ir_fig4} 
\end{figure*}

\begin{figure*}
\flushleft
\includegraphics[width=0.7\textwidth]{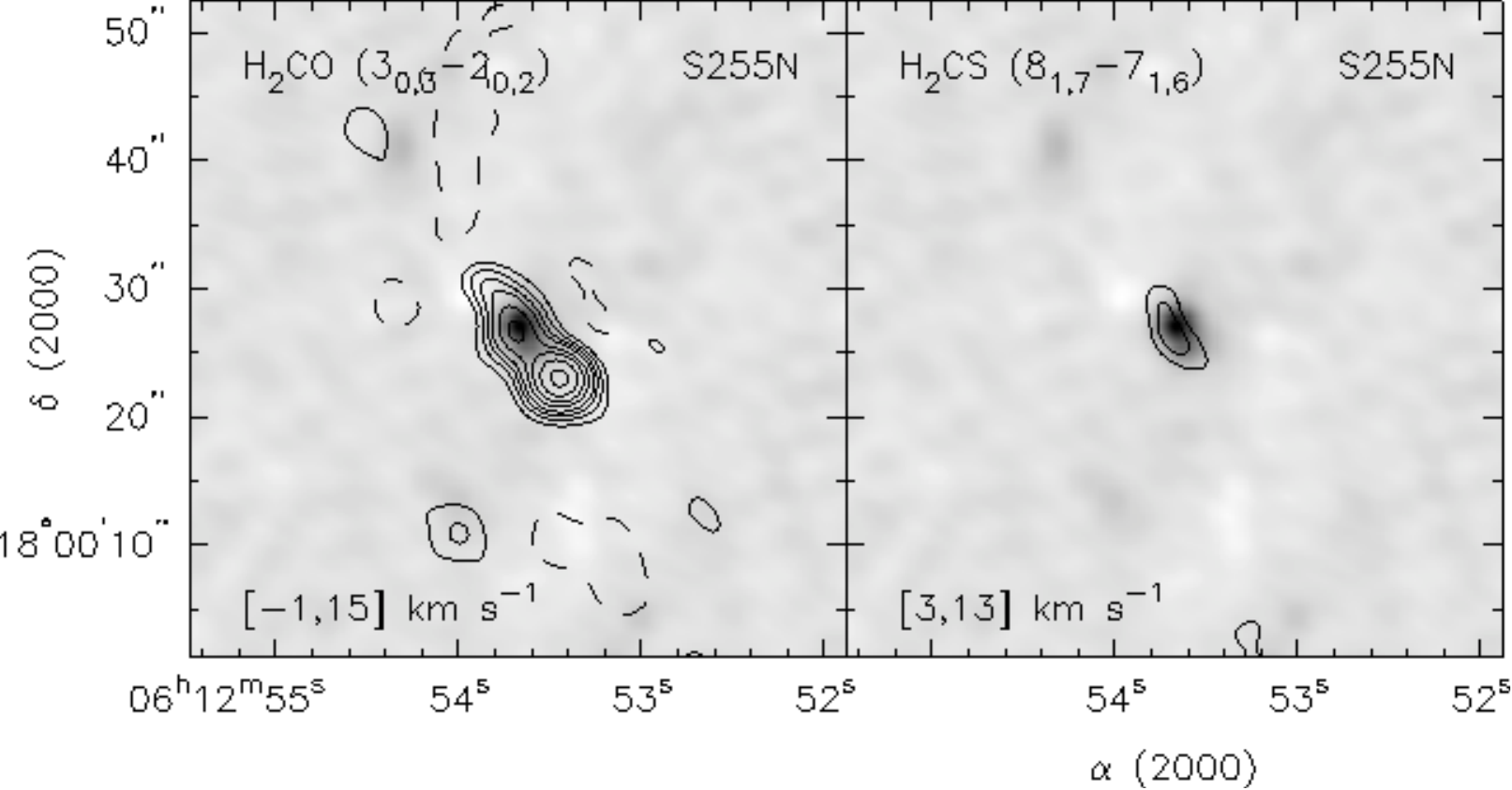} 
\caption{Maps of the H$_2$CO $3_{0,3} -    2_{0,2}$ and H$_2$CS $8_{1,7} -    7_{1,6}$ integrated line emission in the S255N area (contours) overlaid on the 1.1~mm continuum image. The velocity ranges are indicated at the bottom of the plots. The contour levels are (--3, 3, 5, 7, 9, 12, 15, 19, 24)$\times 0.6$~Jy\,beam$^{-1}$\,km\,s$^{-1}$ for H$_2$CO $3_{0,3} -    2_{0,2}$ and (--3, 3, 5)$\times 0.6$~Jy\,beam$^{-1}$\,km\,s$^{-1}$ for H$_2$CS $8_{1,7} -    7_{1,6}$. The dashed contours show negative features due to the missing flux.}\label{fig:n_fig4} 
\end{figure*}

Our data set contains several other important molecular lines including C$^{34}$S, HC$_3$N, SO, SO$_2$, H$_2$CO, H$_2$CS, CH$_3$CN, OCS, and HNCO. Most of them have not been observed with high angular resolution in both these objects before now. Some of these lines, in particular those with relatively high excitation energy (e.g., HC$_3$N and SO$_2$), are detected almost exclusively towards the hot core in S255IR-SMA1. Other molecules, (e.g., SO)  trace the outflows. In Figs.~\ref{fig:ir_fig3},\ref{fig:n_fig3},\ref{fig:ir_fig4},\ref{fig:n_fig4} we present maps of the emission of C$^{34}$S, HC$_3$N, H$_2$CO, H$_2$CS, and SO$_2$. In particular, fairly strong emission is observed in H$_2$CO. It traces most of the new continuum clumps in both S255IR and S255N, and also several other features, including S255IR-N$_2$H$^+$(1). 

It is worth noting that our C$^{34}$S $ J=6-5 $ map of the S255IR area looks very different from the $^{13}$CS $ J=5-4 $ map presented by \citet{Wang11}. Our data show the C$^{34}$S emission is of similar intensity in both S255IR-SMA1 and S255IR-SMA2, and approximately coincident with the dust continuum peaks (Figs.~\ref{fig:spectra},\ref{fig:ir_fig3}).  The $^{13}$CS emission of \citet{Wang11} is seen only in S255IR-SMA1 and is significantly shifted from the continuum peak. Interestingly, our $^{13}$CS observations of the same line show the intensity distribution very similar to C$^{34}$S $ J=6-5 $. The only reasonable explanation is that the bulk of the $^{13}$CS emission is resolved out by the smaller beam of \citet{Wang11} (which is about 2 times narrower than our beam) and they see only a compact structure at the eastern edge of S255IR-SMA1.

\subsection{Decimeter and centimeter wave radio continuum emission}
\begin{figure}
\includegraphics[width=\columnwidth]{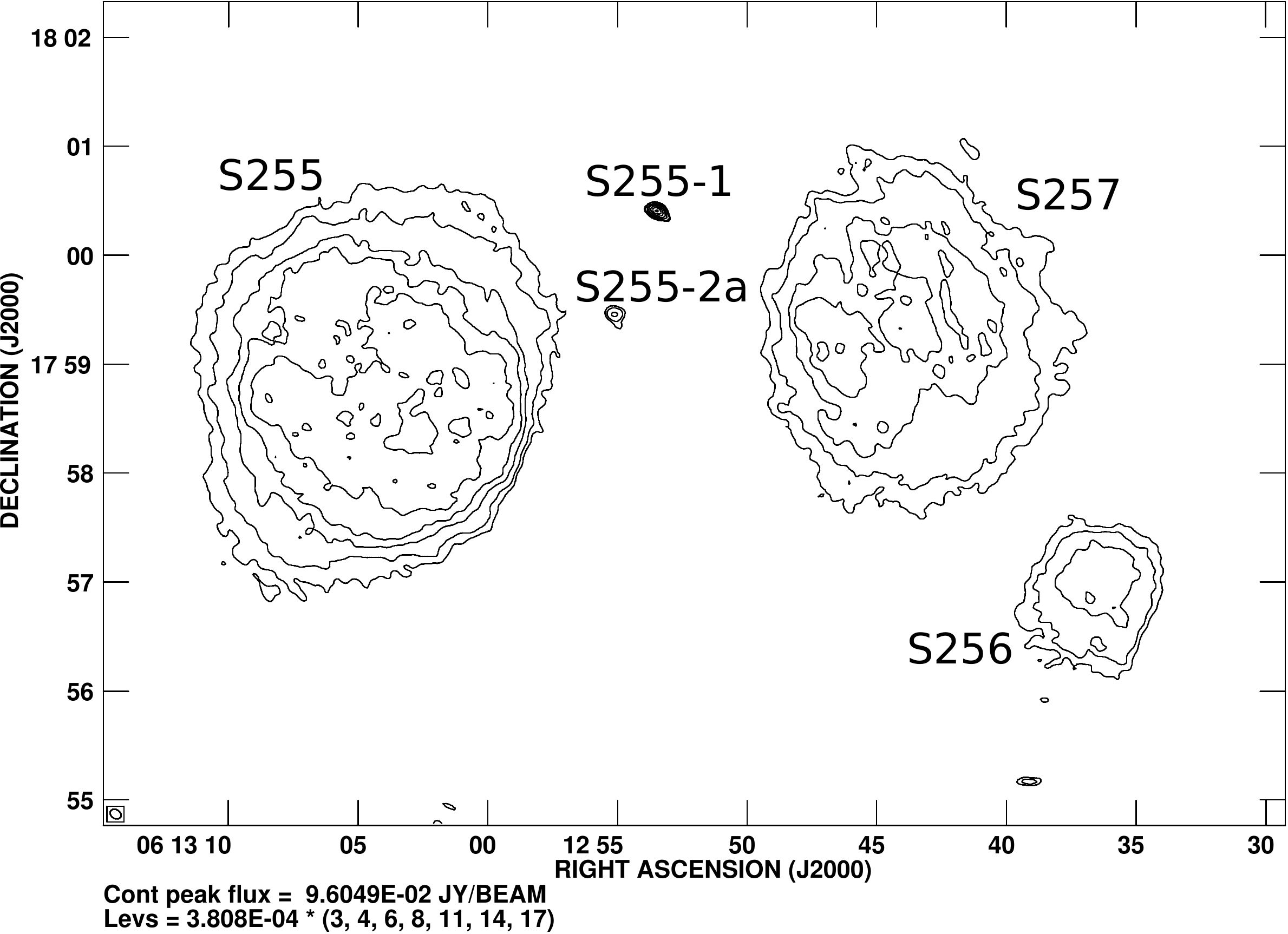} 
\caption{Our GMRT map of the investigated area at 610 MHz. It covers the extended \htwo\ regions S255, S256 and S257. Two compact sources in the center are associated with the S255IR and S255N star forming cores.}\label{fig:610MHz} 
\end{figure}

In Fig.~\ref{fig:610MHz} we show a large-scale 610~MHz map of the region, obtained with the GMRT. It covers the evolved \htwo\ regions S255, S256 and S257. Two compact sources in the center are associated with the S255IR and S255N star-forming cores.

\begin{figure*}
\begin{minipage}{0.48\textwidth}
\centering
\includegraphics[width=\textwidth]{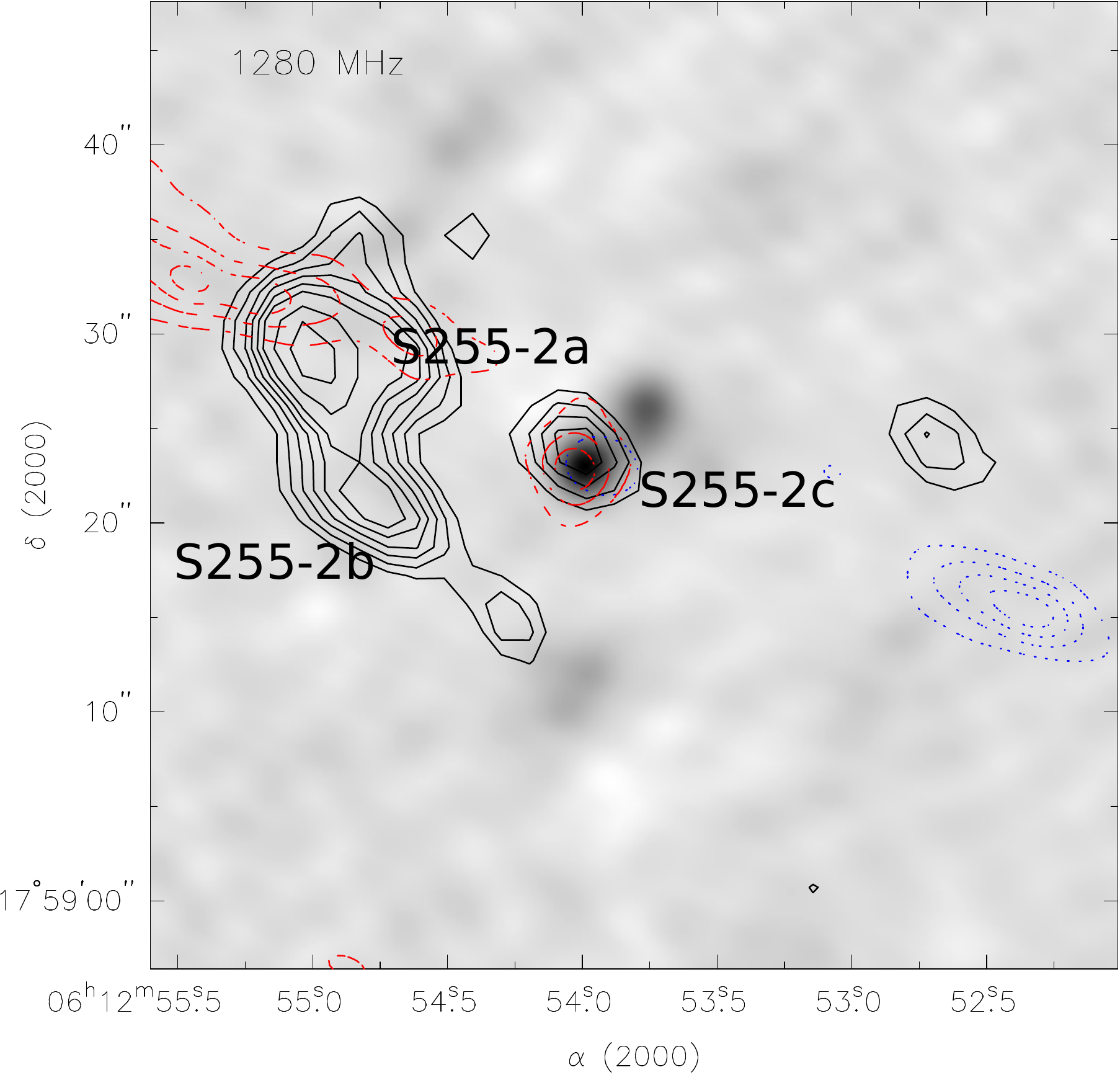}
\end{minipage}
\hfill
\begin{minipage}{0.48\textwidth}
\centering
\includegraphics[width=\textwidth]{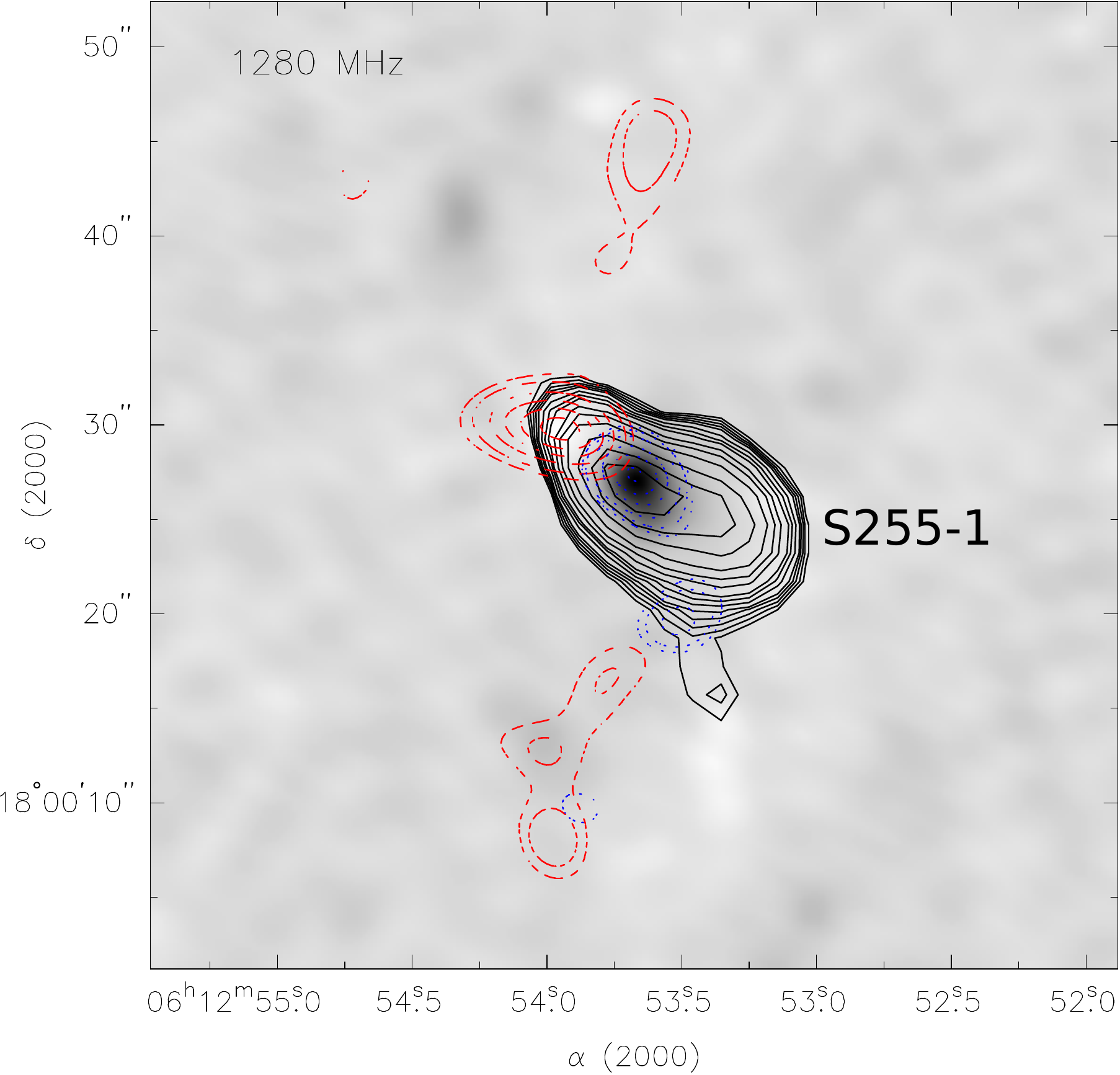}
\end{minipage}
\caption{Maps of the 1280~MHz continuum emission in the S255IR (left panel) and S255N (right panel) areas (contours) overlaid on the 1.1 mm images. The contour levels are (3, 3.5, 4, 4.5, 5, 6, 7, 8, 10)$\times 0.22$~mJy\,beam$^{-1}$ for S255IR and (3, 3.5, 4, 4.5, 5, 6, 7, 8, 10, 12, 15, 20, 25, 30)$\times 0.22$~mJy\,beam$^{-1}$ for S255N. The dash-dotted (red in the color version) and dotted (blue in the color version) contours show the red-shifted and blue-shifted high-velocity CO emission, respectively.}
\label{fig:maps-gmrt}
\end{figure*}

In Fig.~\ref{fig:maps-gmrt} we present 1280~MHz GMRT continuum maps of S255IR and S255N. A comparison of the original GMRT maps with high-frequency VLA archival maps (in particular, at 15~GHz) shows a positional shift of a few arc seconds between the maps. The reasons for this shift are unclear but probably are related to phase calibration at the GMRT. We made positional corrections to the GMRT maps to compensate for the shifts. Thus, the absolute coordinates in these maps are not reliable to more than a few arc seconds.

The main components seen in the GMRT maps were identified by \citet{Snell86} and we follow their nomenclature here for the radio continuum sources (in particular S255-1 corresponds to S255N). Our data extend the frequency range of radio continuum observations and provide important information on the structure of the radio continuum sources at low frequencies. In our 610~MHz map, we only see emission from the sources S255-2a and S255-2b in the S255-IR cluster. These two sources are not resolved in the 610 MHz map, however, they appear as two distinct sources in our higher resolution 1280 MHz map. We measured flux densities and sizes of the radio continuum sources using a two-dimensional elliptical Gaussian model via the AIPS task JMFIT; the results are given in Table~\ref{table:gmrt}. In Section~\ref{sec:disc} we compare our data with observations at other wavelengths and discuss possible relations between the radio continuum and molecular emission.

\begin{deluxetable}{lcccrc}
\tablecaption{Results of the GMRT radio continuum measurements. The fluxes, sizes (convolved) and position angles at 1280~MHz as well as fluxes at 610~MHz are indicated. \label{table:gmrt}}
\tablehead{\colhead{Name} &\colhead{$S_{1280}$} &\colhead{$\theta_{\mathrm{max}}$} &\colhead{$\theta_{\mathrm{min}}$} &\colhead{P.A.} &\colhead{$S_{610}$}\\ \colhead{} &\colhead{(mJy)} &\colhead{$^{\prime\prime}$} &\colhead{$^{\prime\prime}$} &\colhead{($^\circ$)} &\colhead{(mJy)}}
\tablecolumns{6}
\startdata
S255-2a      &3.81 $\pm 0.49$    &8.1  &7.0  &46.4  &4.22 $\pm 0.86$\\
S255-2b      &1.53 $\pm 0.41$    &7.7  &3.8  &55.9 &0.96 $\pm 0.31$\\
S255-2c      &1.11 $\pm 0.32$    &7.7  &4.3  &58.9 &$ <0.4 $\\
S255-1      &20.25 $\pm 0.46$    &10.5  &5.0  &61.1  &10.48 $\pm 0.78$

\enddata

\end{deluxetable}

Although at the VLA only spectral line observations were made we combined line-free channels to form continuum images at 23.7~GHz. These continuum maps are not of very high quality (the usable bandwidth was only about 3~MHz) and so the detection level and uncertainties just are not that good. We detected S255-1 and S255-2c. Their measured fluxes are $19 \pm 2$~mJy and $3.3 \pm 0.5$~mJy, respectively. S255-2a and 2b are not detected. 

\section{Discussion} \label{sec:disc}

\subsection{General morphology and kinematics}
Our data indicate the presence of several new clumps in the mapped areas. Some of them are visible in the 1.1~mm continuum map. Most of these continuum clumps also show molecular emission (although the sets of emitting lines are somewhat different for each one). 
A comparison with our recent NIR J, H and K band observations of this area \citep{Ojha11} shows almost no association of the newly detected continuum clumps with the NIR sources (Fig.~\ref{fig:NIR}). This implies that these clumps are at a very early stage of evolution.

\begin{figure*}
\includegraphics[width=\textwidth]{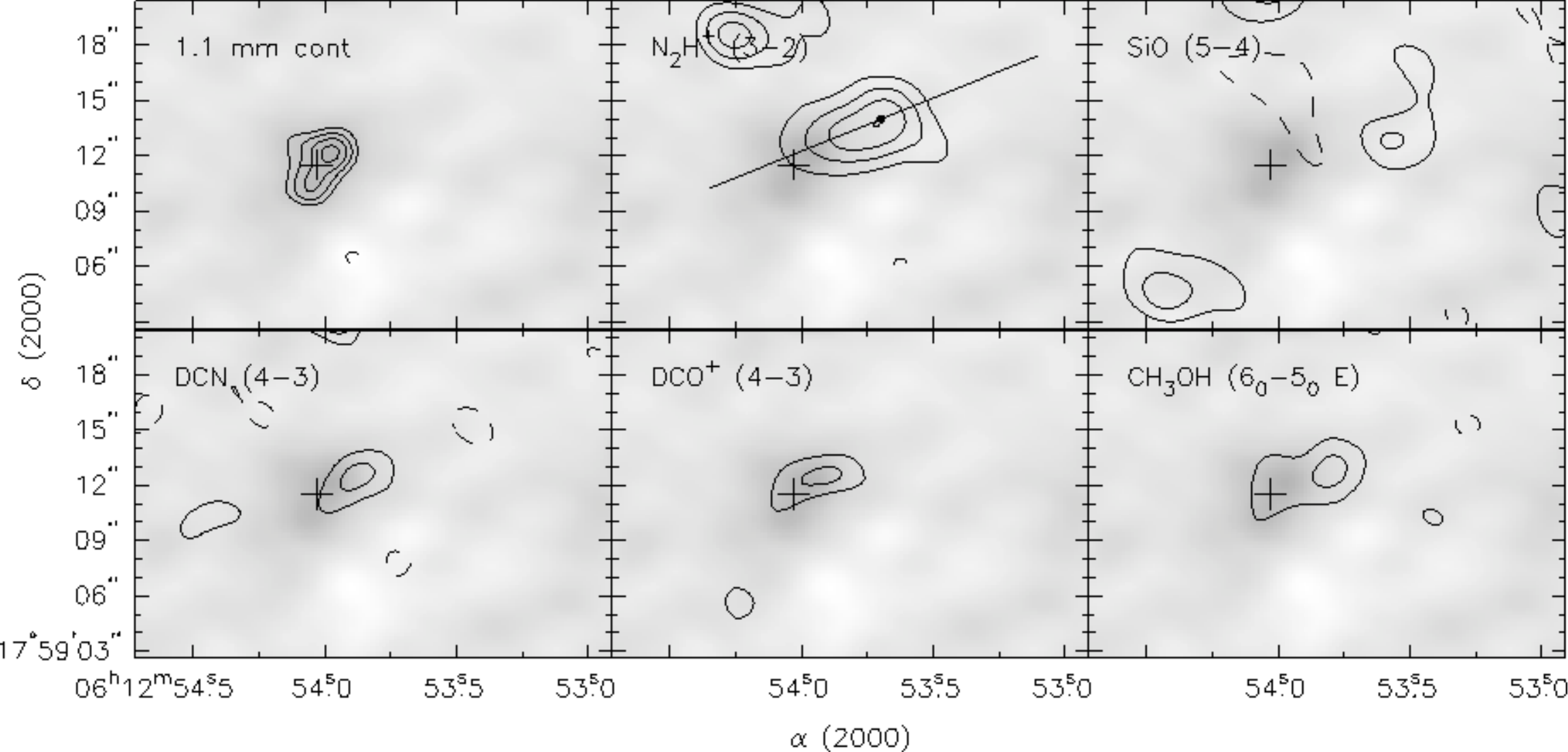} 
\caption{Maps of the 1.1 mm continuum emission and integrated line emission of several molecular species in the S255IR-SMA4 area (contours) overlaid on the 1.1~mm continuum image. The velocity ranges are the same as in Figs~\ref{fig:ir_fig1},\ref{fig:ir_fig2},\ref{fig:ir_fig3},\ref{fig:ir_ch3oh} for those species. The contour levels are the same as earlier for continuum, N$_2$H$^+$ and CH$_3$OH, and 1.5 times lower for SiO, DCN and DCO$^+$ (in order to better show weaker features). The cross marks the position of S255IR-SMA4. The strait line in the N$_2$H$^+$ panel indicates the location of the position-velocity slice shown in Fig.~\ref{fig:ir-mm4_n2h-pv} The dot on this line corresponds to the zero offset in Fig.~\ref{fig:ir-mm4_n2h-pv}. }\label{fig:ir_mm4} 
\end{figure*}

\begin{figure}
\includegraphics[width=\columnwidth]{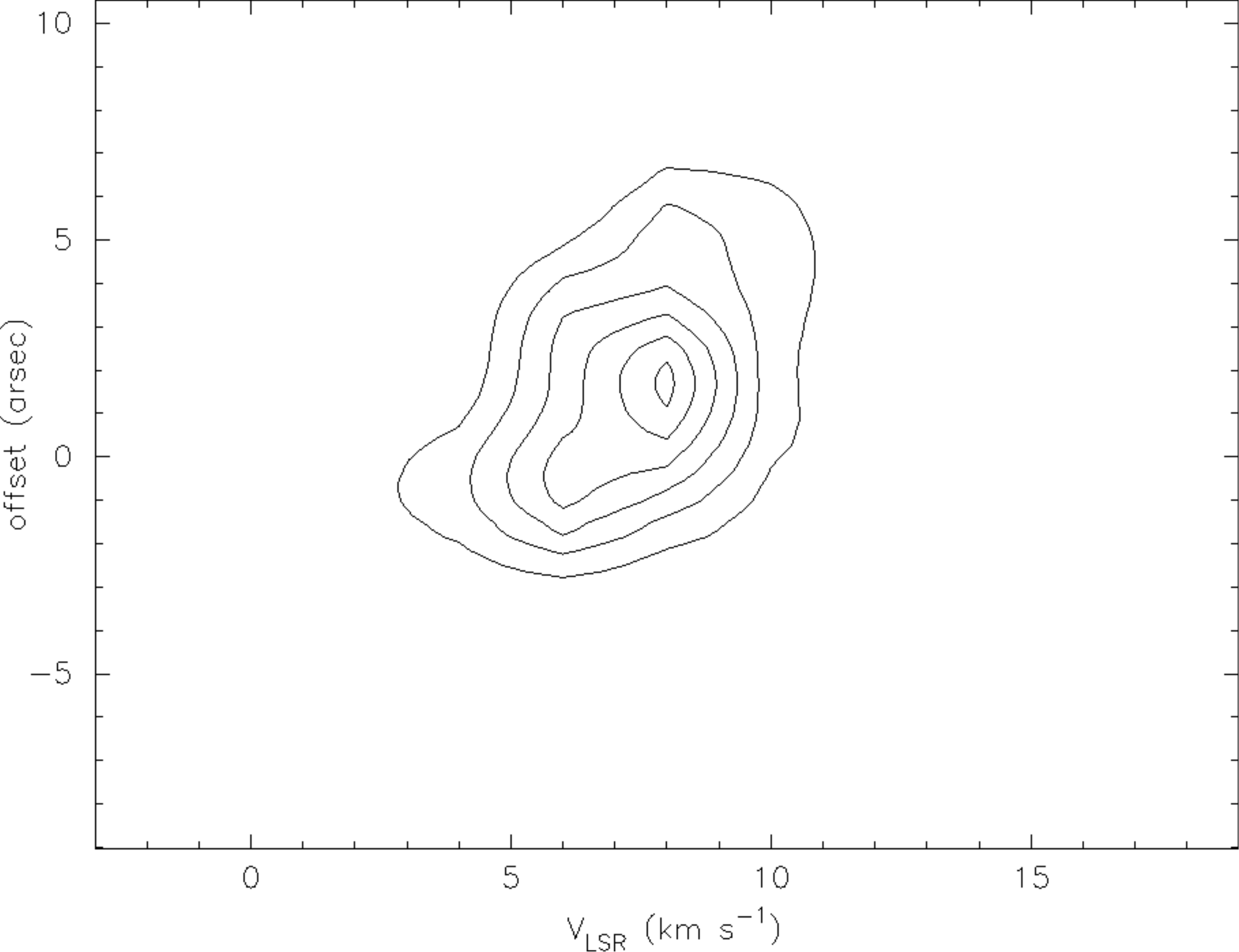} 
\caption{The position-velocity diagram for the N$_2$H$^+$ emission in the S255IR-SMA4 area along the line indicated in Fig.~\ref{fig:ir_mm4}. }\label{fig:ir-mm4_n2h-pv} 
\end{figure}

A pronounced example is the new clump which we designate as SMA4 in the S255IR area. It shows fairly strong continuum emission (Table~\ref{table:cont}) and is associated with emission in many molecular lines including N$_2$H$^+$, DCO$^+$, DCN, H$_2$CO, H$_2$CS, and CH$_3$OH. Interestingly, emission peaks of different species are located at different distances from the continuum source, implying a chemical abundance gradient in approximately the east-northeast direction (Fig.~\ref{fig:ir_mm4}). There is also a moderate velocity gradient in this direction, that can be seen in the N$_2$H$^+$ line channel maps (Fig.~\ref{fig:ir_n2h-chmap}) and is also illustrated by the position-velocity diagram in Fig.~\ref{fig:ir-mm4_n2h-pv}. The total velocity change is about 5~km\,s$^{-1}$. One of the SiO clumps, partly overlapping with the N$_2$H$^+$ emission, also appears to be associated with this object, being located at the largest distance from the continuum source ($ \sim 7'' $) (Figs.~\ref{fig:ir_fig1},\ref{fig:ir_n2h-chmap},\ref{fig:ir_fig3},\ref{fig:ir_sio-chmap}). At the same time the C$^{18}$O emission is very weak. Perhaps we see here an outflow oriented almost perpendicular to the line of sight. This can explain the relatively small velocity gradient and narrow line widths.

The data also show a complex kinematics of some of the previously known clumps. As mentioned above the ammonia lines probably have two velocity components in S255IR-SMA1 and S255N-SMA1. \citet{Wang11} also mentioned that the C$^{18}$O spectrum toward S255IR-SMA1 shows two peaks. The presence of two velocity components is also evident in the spectra shown in Fig.~\ref{fig:spectra}. The most intense emission toward S255IR-SMA1 in various lines including the highest excitation lines, is observed at $V_\mathrm{LSR} \sim 4-5$~km\,s$^{-1}$. Another component, at $V_\mathrm{LSR} \sim 10$~km\,s$^{-1}$ is seen in lines of lower excitation and has a somewhat smaller linewidth. 

In S255N-SMA1 the data also show two kinematic components --- at $V_\mathrm{LSR} \sim 8$ and $\sim 10$~km\,s$^{-1}$. The lines with the excitation energy $ E_\mathrm{u} \ga 40 $~K are observed at $V_\mathrm{LSR} \sim 8$~km\,s$^{-1}$ while lower excitation lines have emission peaks in the range $V_\mathrm{LSR} \sim 8-10$~km\,s$^{-1}$.

\subsection{Basic physical parameters of the clumps}

\subsubsection{Kinetic temperature}
We derive gas kinetic temperature from our ammonia observations using the usual approach described for example in \citet{Mangum92,Harju93}. The NH$_3$ (1,1) and (2,2) spectra were fitted using the GILDAS software package (http://www.iram.fr/IRAMFR/GILDAS). Results of the temperature estimates for the continuum clumps are given in Table~\ref{table:masses}. Typical uncertainties are 20--30\%. As mentioned above, in S255IR-SMA1 and S255N-SMA1 we probably see secondary velocity components. The signal-to-noise ratios for them are very low; we show estimates of the kinetic temperature for these components in parenthesis with question marks. 

\begin{deluxetable*}{lcccccccc}
\tablecaption{Velocities and physical parameters (line width, linear size, kinetic temperature derived from ammonia observations, \textit{assumed} dust temperature, mass, virial mass and mean density) of the millimeter wave continuum sources. \label{table:masses}}
\tablehead{\colhead{Name} &\colhead{$V_\mathrm{LSR}$} &\colhead{$\Delta V$} &\colhead{$L$} &\colhead{$T_{\mathrm{kin}}$(NH$_3$)} &\colhead{$T_{\mathrm{d}}$}\tablenotemark{a} &\colhead{$M$} &\colhead{$M_{\mathrm{vir}}$} &\colhead{$\bar{n}$} \\ \colhead{} &\colhead{(km\,s$^{-1}$)} &\colhead{(km\,s$^{-1}$)} &\colhead{(pc)} &\colhead{(K)} &\colhead{(K)} &\colhead{(M$_{\odot}$)} &\colhead{(M$_{\odot}$)} &\colhead{(cm$^{-3}$)} }
\tablecolumns{9}
\startdata
S255IR-SMA1 &4.4 &3.3 &0.012	&240, (80 ?)	&40    &10 &14 &$ 2\times 10^8 $\\
S255IR-SMA2 &9.2 &2.0 &0.031	&44		&40    &10 &13 &$ 10^7 $\\
S255IR-SMA4 &7.9 &2.6 &0.048	&$ \la 20 $	&20    &14 &34 &$ 4\times 10^6 $\\ 
S255N-SMA1  &8.0 &3.5 &0.030	&29, ($ >200 $~?)	&30    &23 &39 &$ 3\times 10^7 $\\
S255N-SMA2  &8.7 &3.0 &0.028	&  	&20  &2 &26 &$ 4\times 10^6 $\\
S255N-SMA3  &9.0 &$< 2$ &0.037	&29	&30    &2 &$< 10$ &$ 10^6 $\\
S255N-SMA4  &7.0 &4.2 &0.020	&  	&20  &8 &40 &$ 2\times 10^7 $\\
S255N-SMA5  &9.5 &3.9 &0.048	&24	&25    &7 &80 &$2\times 10^6 $\\
S255N-SMA6  &8.9 &4.5 &0.039	& 	&20   &6 &80 &$ 4\times 10^6 $

\enddata
\tablenotetext{a}{assumed}
\end{deluxetable*}

In S255N-NH$_3$ we obtained the temperature of about 13~K for the 8~km\,s$^{-1}$ component and about 23~K for the 10~km\,s$^{-1}$ component. The temperature of the latter component is practically the same in S255N-NH$_3$ and in S255N-SMA1, while the temperature of the 8~km\,s$^{-1}$ component sharply increases toward S255N-SMA1 from  very low values at S255N-NH$_3$. The temperature estimate for S255IR-N$_2$H$^+$(1) is about 50~K.

Our data set contains several methanol transitions of significantly different excitation. In principle they could be used for temperature estimates with the traditional method of rotational diagrams. Unfortunately in most cases the constructed rotational diagrams do not allow any reasonable temperature fit. The reason can be a significant deviation from LTE conditions assumed by the method of rotational diagrams. As mentioned above, several of the observed transitions may be masing. Preliminary results of non-LTE modelling based on the approach described in \citet{Salii06,Sutton04} show a fairly good agreement with the ammonia temperature estimates (S.~Salii \& A.~Sobolev, private communication). Thus we use only these latter estimates here.

\subsubsection{Masses and densities}
Here we present some simple estimates of clump masses, sizes and densities derived from the dust continuum emission. 
Our continuum flux measurements at two different frequencies ($\sim 225$ and $\sim 285$~GHz) imply a spectral index of about 2 for both S255IR-SMA1 and SMA2. This indicates optically thick emission, which is unrealistic.
A comparison of our measured flux densities at 225~GHz for SMA1 and SMA2 with those published by \citet{Wang11} shows that our values are about 2 times higher. This could be caused by different $uv$ ranges of the observations; with our lower resolution data we are less susceptible to missing flux. Our 285~GHz data have an angular resolution closer to that of \citet{Wang11}. Using our 285~GHz flux densities and the 225~GHz flux densities from \citet{Wang11}, we obtain a spectral index of about 4 for these clumps, as expected for optically thin dust emission. This shows that even simple mass estimates based on continuum flux densities are highly uncertain and strongly depend on the $uv$-coverage.

We assume the dust temperatures ($ T_{\mathrm{d}} $) close to the kinetic temperatures derived from the ammonia observations. For the clumps with no ammonia temperatures we assume $ T_{\mathrm{d}}=20 $~K. In S255IR-SMA1 and S255N-SMA1 we see components with different temperatures. For the bulk of the dust emission we assume the lowest temperature. In S255IR-SMA1 the ammonia emission is very weak and estimates of the lowest temperature are very uncertain. We assume $ T_{\mathrm{d}}=40 $~K in accordance with the value of kinetic temperature found in single-dish observations \citep{Zin09}.
Then, we assume a gas-to-dust mass ratio of 100, and adopt a dust absorption efficiency following \citet{Ossenkopf94}. The masses derived in this way are given in Table~\ref{table:masses} and range from 2 to 34~M$_\odot$.

In this table we also indicate velocities and typical line widths for the clumps as well as their linear sizes and virial masses. The velocities and line widths in most cases were derived from C$^{18}$O and/or C$^{34}$S spectra. In some clumps (S255IR-SMA4, S255N-SMA6) these lines are too weak and we used data of other lines, in particular N$_2$H$^+$ for S255N-SMA6 and H$_2$CO for S255IR-SMA4. For S255IR-SMA1 and S255N-SMA1 we give velocities and line widths of the most intense components. 

The clumps are elongated; we indicate linear sizes as the geometric mean between the major and minor axes. The virial masses are obtained in the simplest way, assuming uniform spherical clouds as in \citet{Zin94}.
The same geometry is used to derive mean gas densities. It is worth noting that for all clumps the line widths greatly exceed expected thermal widths which hints at strong turbulence in these objects.

For the clumps investigated by \citet{Wang11} our mass estimates are very close to their estimates. The virial masses for the major clumps in these areas (S255IR-SMA1, S255IR-SMA2 and S255N-SMA1) and for S255IR-SMA4 are close to the masses derived from the dust continuum emission (within a factor of $\sim 2$). This means that these clumps are close to being in gravitational equilibrium. On the other hand, the virial masses of the weaker clumps in the S255N area in some cases are an order of magnitude greater than the masses obtained from the dust emission. In principle, these clumps could in truly be unbound and represent transient density enhancements produced by turbulence. However, estimates of their virial masses are rather uncertain, too. For example, the ammonia lines in S255IR-SMA5 are almost 4 times narrower than the C$^{18}$O line which was used for the estimate of the virial mass. If we estimate the velocity dispersion from ammonia, the discrepancy between the virial mass and mass derived from the dust emission, disappears for this clump. The difference in the line widths can be caused by the fact that the C$^{18}$O emission is much more extended. Then, these clumps could be even colder than we assumed. We can conclude that the question of the stability of these clumps needs a further investigation.  

Our data show clusters of star forming clumps in both S255IR and S255N. Using the data on clump velocities presented in Table~\ref{table:masses} we can estimate virial masses of these clusters. These estimates give $ M_\mathrm{vir} \sim 300 $~M$_\odot$ for S255IR and $ M_\mathrm{vir} \sim 100 $~M$_\odot$ for S255N which is in reasonable agreement with their mass estimates based on single-dish observations of the dust emission \citep{Zin09,Wang11}.

\subsection{Molecular clumps without continuum counterpart}
In some cases, even quite strong molecular emission at velocities of the quiescent gas has no detectable continuum counterpart. For example, the ammonia emission in S255N is strong and extended toward the north in both the (1,1) and (2,2) lines (see Fig.~\ref{fig:n_fig1}).  This emission occurs at distances up to $ \sim 25^{\prime\prime} $ ($\sim 0.3$~pc) from the continuum source. At the same time our own data and the results published by \citet{Wang11} show that the ammonia emission is spatially close to the red-shifted high-velocity CO emission lobe and is extended in the same direction. Moreover, a comparison of the NH$_3$ (1,1) and (2,2) maps shows that the intensity ratio $ I(2,2)/I(1,1) $ increases towards this CO lobe, implying a temperature increase. These facts hint at a physical association between the ammonia emission and the high-velocity outflow, even though they do not coincide spatially or have the same velocity. A possible explanation is that we see in ammonia (and partly in N$_2$H$^+$ and DCO$^+$) the cold quiescent gas interacting with the high-velocity outflow. This interaction heats the gas, causing the observed temperature gradient. The absence of emission in other molecular lines probably means that these molecules are frozen onto dust grains. This assumption is consistent with the low kinetic temperature derived for this area. However, the picture is further complicated by the fact that we see two kinematic components in ammonia (Fig.~\ref{fig:n_nh3}) with significantly different temperatures (Section~\ref{sec:nh3}). An alternative explanation could be that one of the ammonia clumps is the driving source of the high-velocity outflow seen in CO. A point in favor of this interpretation is that the CO emission is not obviously associated with the SMA1 clump, while the ammonia emission is. The interpretation is hampered by the fact that we see no blue-shifted high-velocity CO emission. 

Another interesting zone of molecular emission without associated continuum emission is S255IR-N$_2$H$^+$(1). \citet{Wang11} report methanol emission in the $8_{-1} - 7_{0}$~E transition and suggest that it is related to the shock heating excited by the outflow, and possibly is masing. However, our data show emission from many other lines, including N$_2$H$^+$, H$_2$CO, several methanol transitions, and probably NH$_3$ and DCO$^+$. This seems to rule out the hypothesis of specific excitation of the $8_{-1} - 7_{0}$~E methanol line and indicates the presence of an object with rather unusual properties. It is noteworthy that we detect no C$^{18}$O or C$^{34}$S emission in this area. At the same time the temperature here derived from ammonia, is rather high ($ \sim 50 $~K) which implies a presence of a heating source. This area apparently also needs a further investigation.

\citet{Cyganowski07} and \citet{Wang11} mentioned one more area of a strong molecular emission without a continuum counterpart shifted to southwest from S255N-SMA1 along the direction of the outflow. It is seen also in several our maps (e.g. Figs.~\ref{fig:n_fig3},\ref{fig:n_fig4}).

\subsection{Molecular outflows}
A description of high-velocity outflows in the S255IR and S255N areas based on CO $ J=2-1 $ observations with the SMA was given by \citet{Wang11}. They identified one outflow in the S255IR area and at least one in S255N, mentioning a rather complicated structure for the high-velocity CO emission here, and not excluding the possibility of multiple outflows. Our data --- in particular the SiO observations --- shed new light on this question. First, the SiO data indicate that there are probably outflows originating from the weaker and presumably less-evolved clumps. As mentioned above, we almost certainly see outflows associated with the S255N-SMA3 and S255N-SMA5 clumps. An outflow related to the S255IR-SMA4 clump also seems probable. None of these objects show NIR, radio continuum, or maser emission, which implies a very early evolutionary stage. 

Basic physical properties of outflows are usually derived from observations of CO and --- preferably --- $^{13}$CO and C$^{18}$O \citep[e.g.][]{Cabrit90,Cabrit92,Henning00,Zin02}. In our case we see the high-velocity CO $J=2-1$ emission towards S255N-SMA3 and S255N-SMA5. However neither $^{13}$CO nor C$^{18}$O are detected in the velocity ranges of the SiO outflows. We also do not see an emission in the lines of other potential outflow tracers (e.g. methanol) at these velocities. Therefore we use CO data to estimate physical parameters of these outflows assuming a small optical depth in the CO line wings. Most probably this assumption is incorrect, which means that our estimates of the mass and mechanical parameters represent lower limits. Following the procedures outlined in the mentioned works we derive mass, momentum, energy, size, age, mass loss rate and mechanical force for these outflows (Table~\ref{table:outflows}). The inclination angle was assumed to be 45$^\circ$. The outflows appear to be very young (a few hundred years). The mass loss rate and mechanical force are significantly lower than in S255IR-SMA1 and S255N-SMA1, and comparable to those in S255S \citep{Wang11}.

\begin{deluxetable*}{lccccccc}
\tablecaption{Parameters of the outflows in S255N-SMA3 and S255N-SMA5 (mass, momentum, energy, size, age, mass loss rate and mechanical force). \label{table:outflows}}
\tablehead{\colhead{Name} &\colhead{$M$} &\colhead{$P$} &\colhead{$E$} &\colhead{Size} &\colhead{$t$} &\colhead{$\dot{M}$} &\colhead{$F$} \\ \colhead{} &\colhead{(M$_{\odot}$)} &\colhead{(M$_{\odot}$\,km\,s$^{-1}$)} &\colhead{(erg)} &\colhead{(pc)} &\colhead{(yr)} &\colhead{(M$_{\odot}$\,yr$^{-1}$)} &\colhead{(M$_{\odot}$\,km\,s$^{-1}$\,yr$^{-1}$)} }
\tablecolumns{8}
\startdata
S255N-SMA3  &0.003 &0.15 &$8\times 10^{43}  $	&0.009    &200 &$2\times 10^{-5}  $ &$8\times 10^{-4}  $\\
S255N-SMA5  &0.012 &0.36 &$10^{44}  $	&0.012    &400 &$3\times 10^{-5}  $ &$9\times 10^{-4}  $
\enddata
\end{deluxetable*}

We cannot derive parameters for the possible outflow related to the S255IR-SMA4 clump since we cannot identify the relevant CO emission. 

\subsection{Chemical features}
One of the most remarkable features in our molecular maps is the absence of N$_2$H$^+$ emission from S255IR-SMA1. Any N$_2$H$^+$ emission from this clump is at least an order of magnitude weaker than that from SMA2. This fact cannot be explained by excitation effects and most probably implies a corresponding difference in the N$_2$H$^+$ column density and the N$_2$H$^+$ abundance, taking into account the similar total gas column densities for these clumps \citep{Wang11}. A drop in N$_2$H$^+$ abundance in luminous massive cores was seen in our earlier single-dish observations \citep{Pirogov07,Zin09} and reported recently by others \citep[e.g.,][]{Busquet11,Reiter11,Cortes11}. The usual explanation for N$_2$H$^+$ depletion is its destruction by CO molecules released from dust grains at rising temperature. However, the C$^{18}$O maps do not support this hypothesis. The C$^{18}$O intensities are similar for SMA1 and SMA2, if not stronger towards the latter \citep[][and our data]{Wang11}. 

It is probably not coincidental that the DCO$^+$ distribution follows that of N$_2$H$^+$, as mentioned above. Both species apparently avoid ionized regions. This is also true for S255N and is consistent with our previous suggestion that the primary destruction mechanism for N$_2$H$^+$ in such objects is dissociative recombination due to an enhanced ionization fraction \citep{Pirogov07,Zin09}. At the same time in S255N there is a noticeable difference between the DCO$^+$ and N$_2$H$^+$ maps: there is no DCO$^+$ emission toward the brightest N$_2$H$^+$ peak. In principle this might be caused by temperature distribution in this area --- an enhanced temperature would reduce the abundance of deuterated molecules. It is worth noting that the NH$_3$ (2,2) emission has a peak here while emission peaks of other deuterated species (DCN and DNC) are also shifted from the N$_2$H$^+$ peak. 

Another interesting feature is the rather strong DCN emission from several clumps, including S255IR-SMA1, which is apparently hot, as mentioned above. Although there are observations of rather high DCN abundances in warm (50--75~K), dense ($n \sim 10^7$~cm$^{-3}$) clumps \citep{Leurini06,Parise09} which were successfully explained by chemical models \citep{Roueff07,Parise09} it is hard to expect the same at the temperature of $\sim 150$~K derived for S255IR-SMA1. Recent modelling by \citet{Albertsson11} shows that DCN/HCN abundance ratio sharply drops at temperatures $ \ga 80 $~K.
Possible explanations for a high DCN abundance in this clump might presume a significant amount of relatively cold gas in this clump or a very young age, with insufficient time to change the isotopic ratio. The required time can be less than $ 10^4 $~years as discussed by \citet{Hatchell98}. An inspection of Figs.~\ref{fig:ir_fig2},\ref{fig:ir_fig3} shows that the peak of the DCN emission is noticeably shifted from the continuum peak and SiO emission peak. Most probably DCN traces cold material adjacent to the hot core. The density of this material should be high in order to explain the observed intensity ratio of the DCN $ J=3-2 $ and $ J=4-3 $ transitions. Our temperature estimates from the ammonia observations show a presence of a relatively  cold component in S255IR-SMA1.

In S255N the DCN emission is noticeably stronger than in S255IR; we also detected DNC here. In the framework of time-dependent chemistry \citep[e.g.,][]{Hatchell98} this may indicate that S255N is younger than S255IR. This is also consistent with a lower temperature of S255N in respect to S255IR as found from ammonia observations. However we need estimates of the DCN/HCN abundance ratios.

As mentioned above we did not detect N$_2$D$^+$ in either S255IR or S255N. An upper limit ($ 3\sigma $) on the N$_2$D$^+$ brightness implies a limit on the intensity ratio $ I(\mathrm{N_2D^+})/I(\mathrm{N_2H^+})\la 0.1 $ for the brightest N$_2$H$^+$ emission peaks. In cold dark clouds this ratio can reach the values of $ \sim 0.25 $ but drops to $ \la 0.1 $ at temperatures $ \ga 25 $~K \citep[e.g.][]{Emprechtinger09}. In our case, when temperature estimates for the bright N$_2$H$^+$ emission peaks are available, they show higher temperatures. Therefore, the non-detection of N$_2$D$^+$ is consistent with our temperature estimates.

Some other chemical peculiarities related to molecular emission not associated with a noticeable dust continuum have been mentioned above (in particular the ammonia source S255N-NH$_3$ in the northern part of the S255N area, some of the N$_2$H$^+$ sources, etc.).

\subsection{Properties of the radio continuum sources}
Our GMRT (Figs.~\ref{fig:maps-gmrt}) and available VLA archival data show the presence of several centimeter and decimeter wave continuum sources. The sources S255-2a and S255-2b identified earlier by \citet{Snell86} are associated with NIR objects \citep{Ojha11} that are classified as B-type stars and apparently represent compact \htwo\ regions excited by these stars. It is interesting that we do not see any molecular or dust emission associated with these sources. The high-velocity CO red-shifted lobe is close to S255-2a but it is unclear whether there is a physical association. Possibly the parental cloud material of these stars has  already been dispersed,  indicating a relatively large age.

The source S255-2c is associated with S255IR-SMA1. It has extensions in the directions of both red-shifted and blue-shifted lobes of the molecular outflow, more pronounced at low frequencies. This indicates a gradient in the emission measure. Apparently at least part of the emission is associated with ionized gas in the outflow. \citet{Ojha11} concluded that this source is younger than S255-2a and 2b. This is consistent with our data, which indicate the presence of a hot molecular core.


In S255N the low frequency radio continuum maps also show an extension in the direction of the outflow lobe, as in S255IR. Presumably this also is an indication of emission from ionized, outflowing gas.

\begin{figure}
\includegraphics[width=\columnwidth]{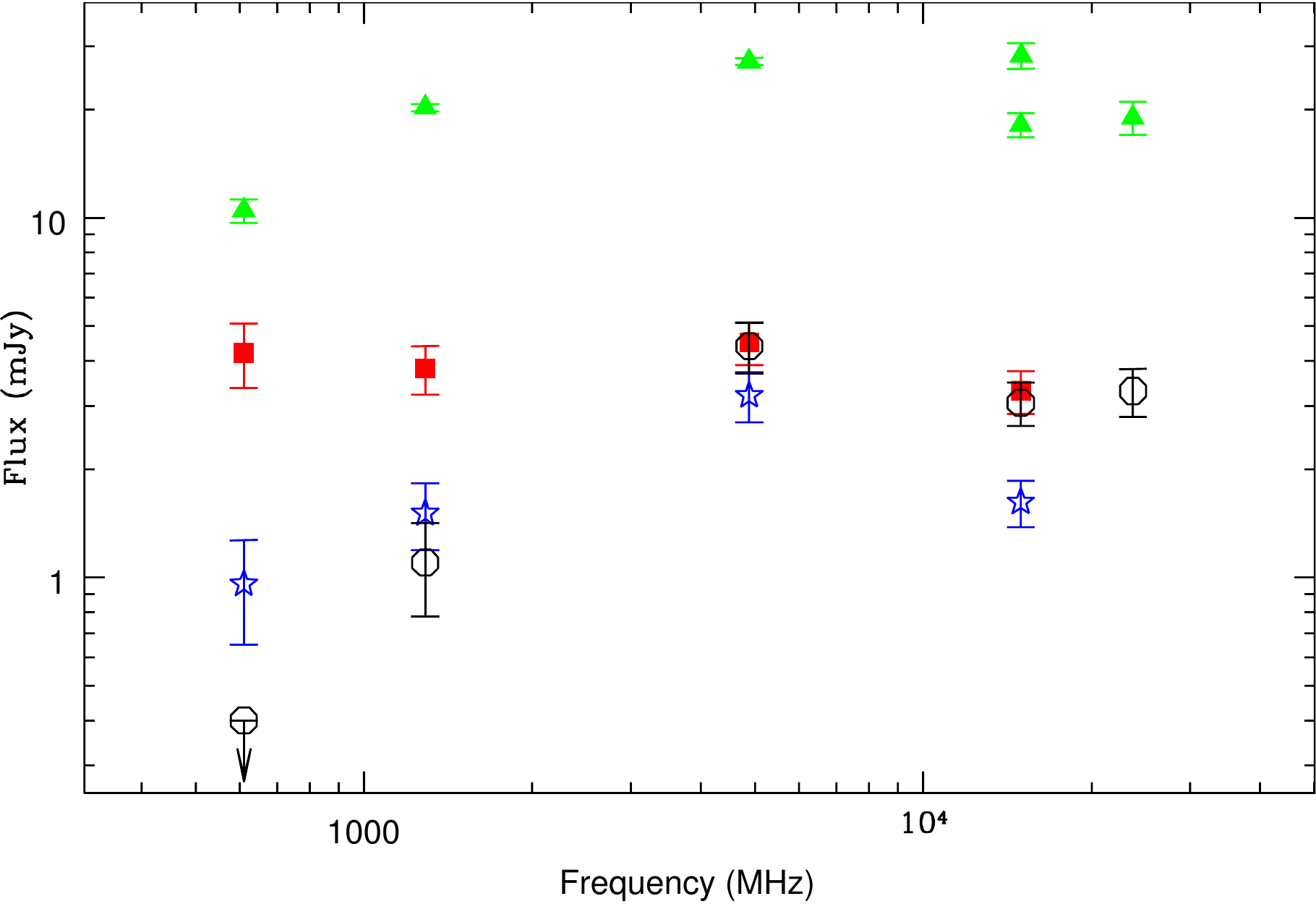} 
\caption{Spectra of S255-1 (triangles), S255-2a (squares), S255-2b (stars) and S255-2c (circles). The data at 610~MHz, 1280~MHz and 23.7~GHz are from the present work, at 5~GHz from \citet{Snell86}, at 15~GHz from \citet{Snell86} (S255-1) and \citet{Ojha11}. }\label{fig:radio} 
\end{figure}

The spectra of all the sources in the range from 610~MHz to 23.7~GHz are plotted in Fig.~\ref{fig:radio} using the
flux densities from the present observations, data from \citet{Snell86} and \citet{Ojha11}. The flux density at 15~GHz is different from that presented by \citet{Ojha11} since the primary beam correction was not applied in that paper. The noise of the map at 610~MHz is used as upper limit for the source S255-2c and the errors at 15~GHz are considered to be 15\% \citep[][private communication]{Ojha11}. It should be noted that these measurements are obtained from radio continuum maps made with different approaches and with different angular resolutions. For example, \citet{Ojha11} estimated positions and flux densities by fitting elliptical Gaussian model components, assuming the sources to be point-like, whereas the total flux density reported by \citet{Snell86} was obtained by integrating  over the region within the 3$\sigma$ limit.  

The structure of S255N is quite complex. The high resolution ($\sim 1$~arcsec) 3.6 cm VLA map by \citet{Cyganowski07} resolved the UC \htwo\ region into three components, with an arc, a point source, and an extended tail, which add uncertainty in the estimation of fluxes. However, the general shape of the SED in Fig.~\ref{fig:radio} is consistent with free-free emission from thermal plasma. 

The spectra show that the emission of S255-2a is optically thin throughout this frequency range, while the other sources become optically thick at the lowest frequencies. An expected flux for S255-2a at 23.7~GHz is about the same as for S255-2c. However S255-2a is more extended (Table~\ref{table:gmrt}) and therefore its brightness should be lower than that of S255-2c. This can be the reason for its non-detection at this frequency. S255-2b is noticeably weaker and apparently below our detection limit at 23.7~GHz.

Using the formulas for free-free emission from \citet{Lang99} and assuming an electron temperature of $10^4$~K we can estimate the emission measure for these objects. For S255-2a with a turnover frequency of $ \nu_\mathrm{T} \lesssim 600 $~MHz, we obtain $EM \lesssim 10^6 $~pc\,cm$^{-6}$. \citet{Ojha11} from their measured fluxes and sizes derived $EM \sim 1.5\times 10^6 $~pc\,cm$^{-6}$ for this source, in agreement with our estimate. For S255-2b, S255-2c and S255-1 the turnover frequency seems to be in the range of 2--3~GHz, which implies $EM \sim (1-2.5)\times 10^7 $~pc\,cm$^{-6}$. For S255-2c and S255-1 these estimates are reasonably close to those derived by \citet{Ojha11}. However, in the case of S255-2b, \citet{Ojha11} obtained a significantly lower estimate for the emission measure which corresponds to the turnover frequency of about 1~GHz. The reasons for this discrepancy are not clear. Perhaps the flux of S255-2b at 5~GHz is overestimated.

\section{Conclusions}
1. We detected several new clumps in the S255IR and S255N areas by their millimeter wave continuum emission. They are also detected in several molecular lines. Temperatures of these clumps found from ammonia observations are $ \sim 20 $~K and their masses are estimated at a few solar masses. These clumps have almost no association with NIR or radio continuum sources, which implies a very early stage of evolution. At the same time, our SiO and other molecular data indicate the presence of high-velocity outflows related to some of these clumps. The outflows associated with S255N-SMA3 and S255N-SMA5 are apparently very young --- a few hundred years only. The line widths of the clumps greatly exceed the expected thermal widths, suggesting significant turbulence in these objects.

2. In some cases there is strong molecular emission at the velocity of the quiescent gas, yet with no detected continuum counterpart. The nature of these sources is not clear. Some of them (e.g., the ammonia source in the northern part of S255N) may represent very cold gas not detectable in the dust continuum. Emission from CO and other molecules would be absent in this case, because they would be frozen onto dust grains. This assumption is consistent with the low kinetic temperature derived for this area from the ammonia observations.

3. N$_2$H$^+$ and DCO$^+$ apparently avoid ionized regions. There is no detectable N$_2$H$^+$ emission associated with the SMA1 clump in S255IR (which contains a hot core and also ionized gas) implying a significant drop in the N$_2$H$^+$ abundance. There is no sign that this is caused by CO evaporation. 

4. We detected rather strong emission in lines of deuterated species --- DCN and DCO$^+$ --- in both S255IR and S255N (in S255N the DCN $ J=3-2 $ line was seen earlier by \citealt{Cyganowski07}). In S255N a DNC emission line is also detected. In S255IR the DCN emission is also quite strong in the vicinity of the hot core. This implies either the presence of a significant amount of cold gas or a very young age for this hot core, insufficient to have changed the isotopic ratio. 

The DCO$^+$ distribution is significantly different from that of DCN, and qualitatively resembles that of N$_2$H$^+$. In particular, both species avoid ionized regions, and both are observed towards the S255N SMA6 clump (in contrast to most other molecules).


5. There is no molecular material associated with the S255-2a and S255-2b radio continuum sources, which are compact \htwo\ regions excited by B type stars. Possibly the parental material of these stars has already been dispersed, indicating a  relatively large age for these objects.


\acknowledgments
We are very grateful to the anonymous referee for the helpful comments and suggestions, and to Andrej Sobolev and Svetlana Salii for discussions of the methanol data.
This work was supported by the Russian Academy of Sciences (Research program No. 17 of the Department of Physical Sciences), Russian Foundation for Basic Research (RFBR), National Science Council (NSC) of Taiwan and Department of Science and Technology (DST) of the Government of India in frameworks of joint research grants (RFBR 08-02-92001-NSC, RFBR 11-02-92690-Ind, NSC 97-2923-M-001-004-MY3, DST-RFBR RUSP-1107).
The research has made use of the SIMBAD database,
operated by CDS, Strasbourg, France.

\bibliographystyle{apj}
\bibliography{apj-jour,s255}

\end{document}